\documentclass[acmlarge, nonacm]{acmart}
\newcommand\app[1]{\textit{SmartKC}}

\AtBeginDocument{%
  \providecommand\BibTeX{{%
    \normalfont B\kern-0.5em{\scshape i\kern-0.25em b}\kern-0.8em\TeX}}}

\acmJournal{IMWUT}
\setcopyright{none}
\renewcommand\footnotetextcopyrightpermission[1]{}
\usepackage[para]{footmisc}
\usepackage{booktabs}
\usepackage{color, colortbl}
\usepackage{array}
\usepackage{float}
\usepackage{enumitem}
\usepackage{soul}
\usepackage{multirow} 
\definecolor{ao(english)}{rgb}{0.0, 0.5, 0.0}

\newcommand{\imwutadd}[1]{\textcolor{black}{#1}}
\newcommand{\imwutremove}[1]{}

\newcommand{\updateremove}[1]{}
\newcommand{\updateadd}[1]{\textcolor{black}{#1}}

\begin{document}

\title{\app{}: Smartphone-based Corneal Topographer for Keratoconus Detection}


 \author{Siddhartha Gairola}
 % Sid OCRID: https://orcid.org/0000-0001-7098-1703
 \email{t-sigai@microsoft.com}
 \orcid{0000-0001-7098-1703}
 \author{Murtuza Bohra}
 \author{Nadeem Shaheer}
 %\email{webmaster@marysville-ohio.com}
 \affiliation{%
   \institution{Microsoft Research}
    \country{India}
 }

 \author{Navya Jayaprakash}
 \author{Pallavi Joshi}
 \author{Anand Balasubramaniam}
 \author{Kaushik Murali}
 \affiliation{%
   \institution{Sankara Eye Hospital}
   \country{India}
 }

\author{Nipun Kwatra}
\orcid{0000-0003-0354-6204}
\email{nkwatra@microsoft.com}
% Nipun OCRID: https://orcid.org/0000-0003-0354-6204
\author{Mohit Jain}
\orcid{0000-0002-7106-164X}
\email{mohja@microsoft.com}
% Mohit OCRID: https://orcid.org/0000-0002-7106-164X
\affiliation{%
  \institution{Microsoft Research}
  \country{India}}

\renewcommand{\shortauthors}{Gairola et al.}


\begin{abstract}
Keratoconus is a severe eye disease affecting the cornea (the clear, dome-shaped outer surface of the eye), causing it to become thin and develop a conical bulge.
% Keratoconus is a severe eye disease which if not treated in time, can lead to near-complete blindness. In many cases this is further accompanied by blurry vision, short-sightedness, irregular astigmatism and sensitivity to light resulting in a poor quality of life. 
The diagnosis of keratoconus requires sophisticated ophthalmic devices which are non-portable and very expensive.
% present in larger (city) hospitals. 
This makes early detection of keratoconus inaccessible to large populations in low- and middle-income countries, making it a leading cause for partial/complete blindness among such populations.
We propose \app{}, a low-cost, smartphone-based keratoconus diagnosis system comprising of a 3D-printed placido's disc attachment, an LED light strip, and a\imwutadd{n intelligent} smartphone app to capture the reflection of the placido rings on the cornea.
An image processing pipeline analyzes the corneal image and uses
the smartphone's camera parameters, the placido rings' 3D location, the pixel location of the reflected placido rings and the setup's working distance to construct the corneal surface, via the Arc-Step method and Zernike polynomials based surface fitting.
In a clinical study with \imwutadd{101} distinct eyes, we found that \app{} achieves a sensitivity of \imwutadd{94.1}\% and a specificity of \imwutadd{100.0}\%. 
Moreover, the quantitative curvature estimates (sim-K) strongly correlate with a gold-standard medical device (Pearson correlation coefficient = \imwutadd{0.78}). Our results indicate that \app{} has the potential to be used as a keratoconus screening tool under real-world medical settings.

% , especially in inaccessible parts of the world and prove useful to physicians as well as users in the diagnosis and tracking of keratoconus disease.
% \textcolor{red}{Keratron (with capital K); Oculyzer (not Pentacam); sim-K1, sim-K2 (not K1, K2); axial, tangential map (not curvature map); keratoconus (full form with small k); small p in placido; use mire and Placido (no rings); use 'non-keratoconus' instead of 'no keratoconus'}
%precision 0.67 and recall 0.86 as compared to the gold standard device Keratron precision 0.70 and recall 1.00.
%We process the captured corneal images with the ring projection to generate axial and curvature heatmaps, along with estimate K1, K2, and PPK values, and evaluate the proposed system for ?? eyes.
%\Sid{Add 4 strong points on (1) What is the existing field, (2) The problem, (3) Our Solution (4) Design implications.}

\end{abstract}



%   <concept>
%       <concept_id>10003120.10003138.10003140</concept_id>
%       <concept_desc>Human-centered computing~Ubiquitous and mobile computing systems and tools</concept_desc>
%       <concept_significance>500</concept_significance>
%       </concept>
%  </ccs2012>

% \ccsdesc[500]{Human-centered computing~Ubiquitous and mobile computing systems and tools}


\maketitle
\section{Introduction} \label{sec:intro}

The cornea is responsible for ${\sim}70\%$ of the refractive power of the human eye~\cite{sinjab_2012}. The shape of the corneal surface plays an important role in determining a person's visual acuity. A typical human cornea is dome-shaped; however, the corneal surface can become distorted due to weakening of the collagen fibers that hold the cornea structure in place. Multiple factors can lead to a distorted cornea, including excessive eye rubbing, genetic abnormalities, overexposure to ultraviolet radiation, and poorly fitted contact lenses. A misshaped cornea can result in multiple medical disorders, such as keratoconus, keratoglobus, and post-LASIK ectasia. In this work, we focus on keratoconus.

Keratoconus is an eye disease characterized by the cornea gradually bulging outward into an irregular conical shape due to progressive thinning and distortion of cornea. 
The cone-shaped curvature weakens the eye's refractive power, resulting in distorted and blurred vision at all distances.
The disease typically affects people between the ages of 10 and 25.
In the United States, 54.5 out of every 100,000 people are diagnosed with keratoconus~\cite{kt-stats-us}; in the Global South, however, the prevalence of keratoconus is much higher.
In India, for example, 2300 out of every 100,000 people suffer from keratoconus~\cite{kt-stats-india}.
This difference can be attributed to a combination of demographics, genetics, and weather conditions.
For example, the Indian population tends to have thinner and steeper corneas~\cite{dharwadkar_2015}, which increases the risk of keratoconus. The hot humid weather in the Global South also contributes towards keratoconus, as it
causes eye irritation leading to frequent eye rubbing.

\begin{figure*}
\begin{center}
    \centering
    \includegraphics[width=1.0\linewidth]{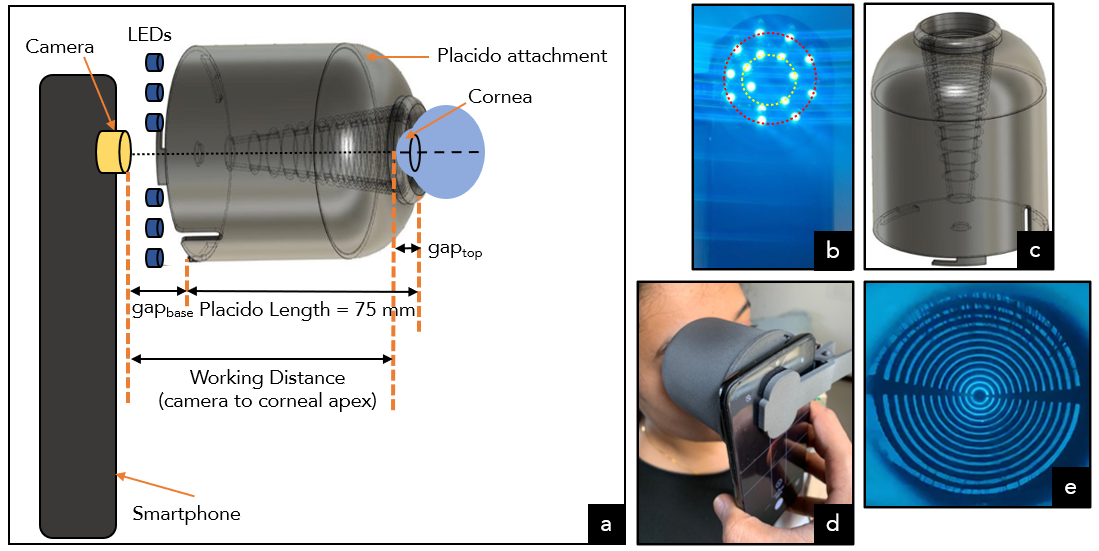}
\end{center}
\vspace{-4mm}
    \caption{\app{} is a low-cost portable smartphone-based corneal topographer: (a) \app{} system setup, (b) LED light arrangement, (c) 3D-printed conical placido clip-on attachment, (d) image capture of a subject's eye, and (e) captured image of the cornea with placido disc reflection (called \textit{mires}).}
    \label{fig:attachment}
\end{figure*}

Corrective lenses can help in treating keratoconous in its mildest form if diagnosed early. In its moderate stages, keratoconus can be mitigated by strengthening the corneal tissues using an effective and low-cost treatment called corneal collagen cross-linking.
In its advanced stages, however, treatment requires an invasive corneal transplant.
In 2012, 27\% of corneal transplants performed worldwide were to treat keratoconus~\cite{stats-cornea-transplant}.
The limited supply of corneas for transplants, the costly surgical procedure, and limited access to eye hospitals has made keratoconus a leading cause of partial and complete blindness among younger demographics in low- and middle-income countries~\cite{kt-epidemiology}.

The clinical gold-standard for diagnosing keratoconus entails mapping the curvature of corneal surface using a technique called corneal topography. Corneal topography is performed via non-invasive imaging devices such as the Optikon Keratron, Medmont E300, Oculus Pentacam, and WaveLight Oculyzer, which use expensive RGB television camera or Schiempflug camera along with a head- and chin-rest.
These devices are expensive, bulky, and require a trained technician to operate. These factors make frequent mass screening of keratoconus hard in remote areas, which is crucial for early detection. A portable, accurate, low-cost diagnostic device can thus significantly help these efforts.

Prior work has proposed low-cost, smartphone-based keratoconus diagnostic solutions~\cite{phone-kt-3D-print, phone-kt-spie-dots, phone-kt-spie-sidepic}.
Most of these solutions leverage a combination of 3D-printed black-and-white concentric rings (called \textit{placido} rings), a macro lens, and an LED circuitry over a smartphone's camera to capture the reflection of the placido rings (called \textit{mires}) on the cornea. In spite of their innovative design solutions, none of them have been satisfactorily evaluated, with the maximum sample size being 11 subjects~\cite{kt-masters-thesis, phone-kt-spie-sidepic, phone-kt-ieee}\imwutadd{. Many require additional hardware \cite{phone-kt-spie-sidepic, phone-kt-ieee, phone-kt-3D-print, phone-kt-spie-dots}, such as chin- and head-rest, tripod and external power source, making them less portable}.
\imwutremove{Moreover, these methods only output a keratoconus classification without any explanation}\imwutadd{Moroever, these methods are based on machine learning classifiers, and output only 2-class/4-class keratoconus classification without any explanation or interpretability}. In contrast, commercial corneal topographers output detailed curvature heatmaps (\textit{e.g.}, axial map, tangential map) along with standard quantitative values (\textit{e.g.}, sim-K1, sim-K2), thus helping doctors to diagnose the presence, \imwutadd{location} and severity of keratoconus. \imwutadd{The heatmaps allow understanding the exact shape of the cornea, which enables explainability and also assists with treatment planning.}

In this work, we propose \app{} --- a low-cost, smartphone-based corneal topographer for keratoconus diagnosis. 
On the hardware side, \app{} comprises of a 3D-printed cone-shaped placido ring attachment (Figure~\ref{fig:attachment}a, c) on a smartphone's camera along with an off-the-shelf USB blue-colored LED light strip (Figure~\ref{fig:attachment}b) to help project concentric black-and-blue rings on the human eye (Figure~\ref{fig:attachment}e).
\app{}'s software analyzes the mires in the captured corneal images to generate assessments that are similar to the clinical gold-standard (the Optikon Keratron, in our case).
The design details of commercial topographers (\textit{e.g.,} placido disc dimensions, camera parameters) are not made public, and their data processing pipeline to generate heatmaps from the captured mire image uses proprietary software. Hence, key contributions of our work is in formulating the mathematical model of the conical placido attachment, \imwutadd{developing a phone app which performs real-time on-device computation for high quality image capture,} and designing algorithms to process mire images for generating curvature heatmaps.
\imwutadd{Developing a portable and low-cost topography system is challenging. We identified several challenges, such as misalignment due to offset/tilt and error in estimation of working distance, and proposed novel algorithmic solutions to address them.} 
% which leads to wrong outputs. To alleviate these we develop a mobile app which provides a realtime feedback to the operator along with automatic image-quality checks to aid capture high quality images.
We evaluate \app{} in a real-world setting without the need for additional infrastructure that is often necessary for commercial topographers, such as chin- and head-rests. 
In our study on \imwutadd{101} distinct eyes (\imwutadd{67} without and \imwutadd{34} with keratoconus) from \imwutadd{57} patients, we found that \app{} achieved a sensitivity of \imwutadd{94.1}\% and specificity of \imwutadd{100.0}\%, close to Keratron's sensitivity of \imwutadd{100.0}\% and specificity of \imwutadd{64.5}\%.
Our system is portable (weighs 140 grams), low-cost (costs \$33 USD), and easy to use.
We also make both the software and hardware details for \app{} publicly available (placido disc attachment .stl file and image processing pipeline source code are available at our \href{https://www.microsoft.com/en-us/research/project/smartkc-a-smartphone-based-corneal-topographer/}{\textcolor{cyan}{project page}}\footnote{\url{https://www.microsoft.com/en-us/research/project/smartkc-a-smartphone-based-corneal-topographer/}}.

The contributions of our work can be summarized as follows:
\begin{itemize}[noitemsep,nolistsep]
    \item An end-to-end implementation of the \app{} system, a low-cost smartphone-based corneal topographer.
    \item A list of identified challenges and novel (hardware and software) solutions\imwutadd{---real-time feedback, on-device image quality check, and auto-click---}to capture high-quality placido reflection (mire) images from a hand-held device.
    \item A suite of techniques \imwutadd{and mathematical formulation} for robust analysis of the captured mire images for qualitative heatmap generation and quantitative simulated-keratometry value computation.
    \item An evaluation of \app{} on \imwutadd{101} distinct eyes from \imwutadd{57} patients in a real-world setting, comparing with a commercial topographer, Optikon Keratron.
\end{itemize}

\imwutadd{To the best of our knowledge, this is the first work that identifies and provides a comprehensive analysis of the challenges posed by a portable device for keratoconus diagnosis in a low-cost setting, proposes novel solutions to alleviate them, and performs an extensive evaluation of the approach in a real-world clinical setting.}
\section{Keratoconus Diagnosis: Corneal Topography}
Accurate measurement of corneal surface curvature, \textit{corneal topography}, is crucial for diagnosing and monitoring progression of several corneal diseases, including keratoconus.
With the advent of high resolution cameras and improvements in technology, a wide variety of commercial topographers have been developed, which can be divided into two categories---curvature-based and elevation-based.
Their working is based on different principles, including placido disc reflection, Schiempflug imaging, and Optical Coherence Tomography (OCT), which have been discussed in the corneal topography surveys~\cite{about-kt-1, about-kt-2, dharwadkar_2015, about-kt-4, about-kt-5}. Here, we briefly discuss them, to understand the difference between these methods and highlight the need for a low-cost portable device for accurately modeling the topography of cornea.

\noindent
\subsection{Curvature-based Topographers}
These topographers measure the anterior (outer) surface of the cornea by projecting light patterns onto the cornea and analyzing the resulting reflection, either directly (by an observer's eye) or via a camera attached to a computer.

\subsubsection{Keratometry}
A keratometer is a simple instrument based on the reflection of light, which provides a quantitative measurement of corneal curvature. The doctor projects a bright white ring at the center of the patient's eye, and the size of its reflection is used to measure the corneal radius in the central 3mm region of the cornea. Though this approach is simple, 
keratometers only measure the central region (${\sim}6\%$) of the anterior corneal surface~\cite{sinjab_2012}).

\subsubsection{Photokeratoscopy}
A photokeratoscope (or keratoscope) projects light on a placido disc target to measure ${\sim}70\%$ of the anterior corneal region~\cite{sinjab_2012}. The placido disc is a series of concentric black and white rings.
In a photokeratoscope, the placido disc is on a back-lit flat (2D) surface, with a central hole for the observer's eye.
It is positioned in front of the cornea and the reflected rings are analysed qualitatively by the observer/doctor to estimate the shape of the cornea. Small, narrow and closely spaced rings indicate steep regions with a small radius of curvature, whereas rings that are far apart suggest flatter regions with a large radius of curvature~\cite{sinjab_2012}. The major limitations of a photokeratoscope device is that it only provides a qualitative assessment of the cornea, and its accuracy is susceptible to de-focusing, misalignment, and tear film disturbances.
Moreover, these devices do not capture/store an image of the eye to be referred to later.

\subsubsection{Videokeratoscopes}
These are modern topographers which provide both a qualitative and a quantitative assessment of the cornea.
The device comprises of a three-dimensional placido disc in a bowl/cone shape illuminated by a light source, and it captures the rings' reflection on the cornea using a camera.
These concentric reflection of the placido rings on the captured image are called \textit{mires}.
Videokeratoscopes are connected to a computer to store and analyze the image.
A mathematical software on the computer is used to compute the distance between the mires sampled on a large number of points (7168 points by Optikon Keratron) to reconstruct the corneal curvature. The output of such devices are different corneal topography heatmaps (like axial map, tangential map), and a set of quantitative metrics, like percentage probability of keratoconus (PPK), degree of astigmatism, sim-K1, sim-K2, \textit{etc}.~\cite{CLMI, about-kt-1, about-kt-4}.
The placido reflection image, topography heatmaps and quantitative values are stored digitally in the computer, to be referred later for tracking the progression of keratoconus.

The placido size, the number of rings, and the color of backlight, vary across topographers from different manufacturers. A larger placido bowl/cone allows more rings to be included, to enable estimating a wider corneal region. 
For the measurement from videokeratography to be valid and reproducible, a few manual adjustments are required, such as appropriate distance between the eye and the device, alignment between the camera center and the cornea center, etc.~\cite{sinjab_2012}, making the role of the device operator crucial.
There are various commercially available devices like Optikon Keratron\footnote{\url{https://www.optikon.it/index/products-details/l/en/p/corneal-keratron}}, Medmont E300\footnote{\url{https://medmont.com.au/e300-corneal-topographer/}}. They have a high degree of precision and accuracy for measuring corneal surfaces~\cite{accuracy_videokt}, high repeatability of measurements~\cite{performance_videokt}, and perform well for early diagnosis of keratoconus~\cite{videokeratography-kt}.

\noindent
\subsection{Elevation-based Topographers}
Placido-based videokeratoscopes are limited to providing assessment of only the anterior surface of the cornea. Without any posterior (inner) surface information, the pachymetric evaluation ( \textit{i.e.}, corneal thickness measurement) is not possible. A pachymetric map is required to perform any modern refractive surgery~\cite{sinjab_2012}. Thus, information about both the anterior and posterior surface is important in the holistic assessment of the cornea. 
The elevation-based devices are of three types: Schiempflug imaging, slit-scanning and OCT-based.
Here, we only describe the Schiempflug imaging devices as they are the most commonly used elevation-based topographers.

\subsubsection{Schiempflug Imaging Devices}
These devices (such as Oculus Pentacam and WaveLight Oculyzer\footnote{\url{https://www.pentacam.com/int} and \url{http://www.eyeclinic.com.gr/en/eksoplismos/item/26-allegro-oculyzer-wavelight.html}}) use a rotating camera to acquire cross-sectional images of the cornea. The camera is based on the Schiempflug imaging principle~\cite{sinjab_2012, schiempflug_principle}, to increase the depth of focus, thus enabling imaging from the anterior corneal surface to the posterior lens surface. The captured images are then used to generate an accurate 3D reconstruction of the cornea. Hence, apart from the output of videokeratoscopes, a Schiempflug imaging device calculates and generates corneal posterior (axial and tangential) maps, pachymetry (cornea thickness) map, and a more accurate elevation map~\cite{dharwadkar_2015}.

Although, devices that can map both anterior and posterior surface topography provide a more detailed corneal assessment, for keratoconus diagnosis, anterior surface topography is sufficient. Only rare conditions, such as `posterior keratoconus'~\cite{posterior-kt} which result in thinning of only the inner surface of the cornea, need a posterior topographer for diagnosis. In our work, we focus on anterior surface topography which is adequate for initial diagnosis of most keratoconus cases. Our proposed system, \app{}, is similar to a videokeratoscope, uses a back-illuminated placido disc, and outputs both quantitative values and qualitative heatmaps. We compare the output from our system against the gold-standard Optikon Keratron videokeratoscope.
\section{Related Work}
Smartphone-based health diagnostic and monitoring solutions have recently gained a lot of attention due to their ubiquity, affordability, convenience, and ease of use.
The in-built smartphone sensors have been exploited for a variety of mobile health applications, such as monitoring jaundice in newborns using images captured from a smartphone~\cite{bilicam}, measuring lung function using audio-recorded by the phone's microphone~\cite{spirosmart}, and estimating blood pressure using accelerometer and camera data~\cite{seismo}. Here, we focus on eye-related mobile health solutions, followed by phone-based keratoconus diagnosis applications.

\subsection{Smartphone-based Ocular Diagnostic Solutions}
Several phone-based eye-related diagnostic applications have been proposed~\cite{netra, catra, pupilscreen}. One of the earliest work, NETRA~\cite{netra} estimate refractive errors in the eye using a microlens array over the smartphone display and asking the user to interactively align patterns seen from a pinhole plane on the display.
Following the same principle, NETRA Autorefractor\footnote{\url{https://eyenetra.com/}} is a self-test refraction tool available for \$1290, to measure refractive error (spherical, cylindrical, and axis) through a series of game-like interactions in a virtual-reality environment.
Other solutions available from NETRA are Netrometer to measure the refractive power of a lens, and Netropter is a phone-based phoropter to measure visual acuity by testing individual lenses on each eye.
CATRA~\cite{catra} provides an interactive way to assess cataracts in the human eye, and EyeMITRA~\cite{eyemitra} is a wearable camera to perform retinal imaging.
Similarly, D-EYE\footnote{\url{https://www.d-eyecare.com/}} is a commercially-available smartphone lens attachment available for \$435 to perform retinal imaging.
These systems along with a smartphone use additional optical and/or electronic system, leading to increased cost.
A recent project, PupilScreen~\cite{pupilscreen} uses a 3D-printed box (similar to Google cardboard) to stimulate the patient's eyes using the phone's flash and records the response using the camera.
The recorded video is analyzed to measure the patient's pupillary light reflex, which is used to assess traumatic brain injury.
Similar to PupilScreen, our proposed phone-based keratoconus diagnosis system has no additional optical hardware, and uses a 3D-printed concentric ring attachment along with a commercial off-the-shelf LED array, thus minimizing cost and maintenance.

\begin{table}[]
\centering
\caption{\imwutadd{Related work summary, including approaches existing in prior literature (rows 1-6) and modern commercial corneal topographers (rows 7-8) in contrast to our proposed device \app{} (row 9).}}
\resizebox{\textwidth}{!}{
\imwutadd{
\begin{tabular}{|l|l|l|l|l|l|l|l|l|}
\hline
\multicolumn{1}{|c|}{\textbf{Prior Work}} & \multicolumn{1}{c|}{\textbf{Coverage}} & \multicolumn{1}{c|}{\textbf{Mires}} &
\multicolumn{1}{c|}{\textbf{Points}} &
\multicolumn{1}{c|}{\textbf{Portable}} & \multicolumn{1}{c|}{\textbf{Extra Hardware}} & \multicolumn{1}{c|}{\textbf{Evaluation}} & \multicolumn{1}{c|}{\textbf{Dataset}} & \multicolumn{1}{c|}{\textbf{Outputs}} \\ \hline
\citet{phone-kt-spie-sidepic} & - & - & - & Yes & - & Hospital & \begin{tabular}[c]{@{}l@{}}10 people / 20 eyes\\ 6 KC / 14 Non-KC\end{tabular} & 4-class classification \\ \hline
\citet{phone-kt-ieee} & - & - & - & No & Chin-head rest & - & 10 people / 20 eyes & 4-class classification \\ \hline
\citet{phone-kt-spie-dots} & $\sim$80\% & 20 & \textless 2000 & No & \begin{tabular}[c]{@{}l@{}}LEDs, Tripod, External power, 3D attachment\end{tabular} & - & - & - \\ \hline
\citet{kt-masters-thesis} & - & 14 & - & No & \begin{tabular}[c]{@{}l@{}}LEDs, Macro lens, External power\end{tabular} & Lab & \begin{tabular}[c]{@{}l@{}}11 people / 22 eyes\\ 3 KC / 19 Non-KC\end{tabular} & \begin{tabular}[c]{@{}l@{}}2-class classification,\\Zernike coefficients\end{tabular} \\ \hline
\citet{phone-kt-3D-print} & - & 20 & - & Yes & \begin{tabular}[c]{@{}l@{}}LEDs, External battery, 3D attachment\end{tabular} & - & - & Heatmaps \\ \hline
Bull's Eye\cite{bullseye} & - & 14 & - & Yes & LEDs, Macro lens, 3D attachment & - & - & - \\ \hline \hline
Optikon   Keratron & $\sim$90\% & 28 & 7168 & No & \begin{tabular}[c]{@{}l@{}}Chin-head rest, Foot-switch, Computer\end{tabular} & Hospital & - & \begin{tabular}[c]{@{}l@{}}PPK, Sim-K, Heatmaps\end{tabular} \\ \hline
Medmont   E300 & $\sim$90\% & 32 & 9600 & No & \begin{tabular}[c]{@{}l@{}}Chin-head rest, Computer\end{tabular} & Hospital & - & \begin{tabular}[c]{@{}l@{}}Sim-K, Heatmaps\end{tabular} \\ \hline \hline
SmartKC (ours) & $\sim$75\% & 28 & 10080 & Yes & \begin{tabular}[c]{@{}l@{}}LEDs, 3D attachment\end{tabular} & Hospital & \begin{tabular}[c]{@{}l@{}}57 people / 101 eyes\\ 34 KC / 67 Non-KC\end{tabular} & \begin{tabular}[c]{@{}l@{}}Sim-K, Heatmaps\end{tabular} \\ \hline
\end{tabular}
}
}

\label{tab:related-summary}
%\vspace{-8mm}
\end{table}
\subsection{Smartphone-based Keratoconus Diagnostics}
Recently, low-cost, portable, smartphone-based keratoconus screening and diagnostic solutions have been proposed~\cite{phone-kt-spie-dots, phone-kt-spie-sidepic, phone-kt-3D-print, phone-kt-ieee, lvpei-phone-kt, bullseye, kt-masters-thesis}.
Most such solutions~\cite{phone-kt-3D-print, phone-kt-spie-dots, lvpei-phone-kt} rely on a 3D-printed placido disc attachment, attached to the smartphone's camera to project concentric rings over the surface of the examined cornea.
The smartphone camera is used to capture an image of the cornea with the placido ring reflection. The captured mire image is then processed for keratoconus screening.
Pinheiro \textit{et al}.~\cite{phone-kt-3D-print} designed a smartphone attachment comprising of a conical placido, an optical system for magnification, a printed circuit board with LEDs, battery and switch, and a phone cover disguised enclosure for fitting the electronic, optical and placido disc parts.
Similar placido disc based design with a combination of LED illumination and magnification lens has been adapted by~\cite{lvpei-phone-kt, bullseye, kt-masters-thesis}.
In \cite{kt-masters-thesis}, 14 concentric rings were used along with 36 3V LEDs in a 6x6 grid and an external lens of 45-mm fixed focal length.
This setup was evaluated with 11 participants (8 normal and 3 keratoconus), and obtained high false positive rate---3 out of 8 normal were classified as having keratoconus.
To increase data points per image, Garcia \textit{et al.}~\cite{phone-kt-spie-dots} use an array of circular holes, instead of the placido-based design, such that each hole resulted in a circular spot of size 0.006mm radius on the image plane. \imwutadd{However, the segmentation and detection of these circular spots in the reflected image is error-prone \cite{phone-kt-spie-dots2}}. In spite of the detailed system description, most of these works~\cite{phone-kt-3D-print, phone-kt-spie-dots, lvpei-phone-kt, bullseye} were not evaluated with any keratoconus patient.
In contrast, \app{} \imwutremove{has 28 rings, with 360 points per mire, and }was evaluated with \imwutadd{101} distinct eyes \imwutadd{(67 non-keratoconus and 34 keratoconus)}  in a clinical setting. \imwutadd{Moreover, \app{} provides high coverage with 28 rings and 360 points per mire, resulting in 10080 points per image.}

Another approach relies on the conical corneal shape during keratoconus~\cite{phone-kt-spie-sidepic, phone-kt-ieee}, and capture the side-view of the cornea (at a 90-degree angle) without any additional attachment.
The captured images are processed using edge detection followed by slope computation~\cite{phone-kt-spie-sidepic}.
They used a dataset of 20 eyes (14 normal and 6 keratoconus), and achieved an accuracy of 87\% for keratoconus and healthy eye classification\imwutadd{, but do not specify the train-test split}.
In a follow-up work, Askarian \textit{et al}.~\cite{phone-kt-ieee} built a chin- and head-rest fixture to capture panoramic 180-degree images of the eye, and enhanced the edge detection algorithm, followed by using SVM for classification. This helped them increase the average accuracy to 89\%~\cite{phone-kt-ieee} with 20 eyes data\imwutadd{, with a split of 70\% train and 30\% test images,\textit{ i.e.}, the system was evaluated on only 6 images.} Also, the chin- and head-rest setup made the system non-mobile.

To summarize, a majority of the proposed low-cost smartphone-based keratoconus diagnosis solutions have not been evaluated, and others have been evaluated with a small sample size of $\sim$10 participants.
Moreover, all these approaches provide 2-class (keratoconus vs no keratoconus) or 4-class (no, moderate, advanced, or severe keratoconus) classifications as output, while the gold-standard medical topographers (like Optikon Keratron, Medmont E300, \textit{etc}.) generate a variety of corneal heatmaps (including axial map, tangential map, elevation map), and provide standard quantitative values (like sim-K1, sim-K2, PPK, \textit{etc}.), which doctors use to not only diagnose, but also understand the severity, location and progression of keratoconus \imwutadd{(refer to Table \ref{tab:related-summary} to get a summary of the key differences between the approaches existing in prior literature, commercial topographers and our proposed system \app{})}.
In this work, we propose a low-cost, smartphone-based keratoconus diagnostic system, using a 3D-printed placido attachment, an off-the-shelf USB LED light strip\imwutadd{, and an intelligent data capture smartphone app}.
We process the captured mire images to generate axial and tangential heatmaps, along with estimated sim-K1 and sim-K2 values.

\section{SmartKC System Design}
\label{system}

Our system is based on the principle behind placido-based corneal topographers such as Optikon Keratron and Medmont E300. In these systems a placido disc pattern from an LED-illuminated conical or ellipsoidal head is projected on the cornea. The projection forms circular mires on the cornea, an image of which is taken via a camera attached to the center of the projection head. For a healthy dome-shaped cornea, the mires will be perfectly circular. Also, the spacing between consecutive mires is a function of the curvature. A distortion in the corneal surface will result in mires coming closer or farther depending on the local curvature---closer mires indicate higher curvature (\textit{i.e.}, small radius of curvature), whereas mires farther apart indicate flatter regions (see Figure~\ref{fig:challenges} III-b,c with steep and flat regions represented in red and blue colors, respectively). 
Therefore, deformities in the corneal surface at different locations can be estimated by analysing the degree of deformation of the mires.
Four sets of data points---camera parameters, 3D location of the placido rings with respect to the cornea, pixel location of the mires on the cornea, and working distance---are combined, using the Arc-Step method~\cite{arc-step} and Zernike polynomials based surface fitting~\cite{zernike_1, zernike_2, zernike_3}, to construct the anterior corneal surface. The curvatures at any point on this surface can then be computed by standard differential geometry, thus generating the axial and tangential maps~\cite{tan_to_axial_klein, arizona_curvature} (Section~\ref{sec:heatmaps}).

\app{} uses the same principle and employs a 3D-printed placido attachment and an off-the-shelf LED array (Figure~\ref{fig:attachment}a), to project a ring pattern on the cornea. The mire image is captured by the smartphone camera, which is then processed by our analysis pipeline (Figure~\ref{fig:process_pipe}) to generate the axial and tangential topography maps. Here, we describe the components of our proposed system, focusing on the challenges faced in developing a portable topographer.

\subsection{Placido Attachment}\label{sec:placido_attachment}
We built a 3D-printed attachment to project the placido ring pattern on the cornea (Figure~\ref{fig:attachment}a). Structurally, it is a conical head comprising of an alternating black and empty ring distribution, with two supports at 0$^{\circ}$ and 180$^{\circ}$, covered with a capsule shaped enclosure. To ensure enough light passes through the empty ring regions, we added an illumination layer consisting of a set of circular blue-colored LEDs. This results in the formation of a concentric alternating blue-and-black ring pattern on the cornea. A small hole is located at the center of the cone aligning with the phone's camera.

\subsubsection{Conical Head}
Commercial topographers have different projection heads---planar, ellipsoid and conical~\cite{sinjab_2012}. Planar heads can accommodate fewer rings, limiting the measuring area to only the central cornea region. Ellipsoidal heads accommodate more rings, thus can measure a larger corneal region with high accuracy, but are bulky (\textit{e.g.}, Zeiss Atlas 9000). To meet our requirements of a portable and robust setup, a conical head is ideal. Conical heads can accommodate 28-32 rings in a compact manner~\cite{placido_design}, thus offering high coverage and high accuracy. They are used in commercial topographers like Optikon Keratron and Medmont E300. Our conical head has 28 rings, a cone length of 70mm, smallest ring (closest to camera) radius of 4mm, largest ring radius of 15mm, and cone semi-vertical angle of 8.93$^{\circ}$ (Figure~\ref{fig:attachment}a). The thickness of the rings is 1mm. Using the mathematical model described in \cite{placido_design}, we designed the placement between placido rings such that the mires formed on a sphere of radius 7.8mm (which is the average radius of a human cornea) are of equal thickness. Our final setup has high precision in a measurement region of $\sim$8.7mm diameter on the cornea (the average diameter of the corneal anterior surface is 11.7$\pm$0.42mm), covering $\sim$75\% of the cornea.

\subsubsection{Support}
In an image captured using a commercial topographer (like Keratron), the consecutive black-and-white mires are formed without any break, as the black part of the placido disc is made up of an opaque material, while the white part is made up of a transparent plastic/glass material, allowing the illuminating light to pass.
However, we found 3D-printing a multicolor cone, comprising of black opaque and transparent alternating rings, to be very expensive (\$310).
For our low-cost setting, we opted for a support-based design, wherein the black and empty ring distribution were connected by two 2mm wide supports at 0$^{\circ}$ and 180$^{\circ}$.
However, these supports resulted in the formation of two v-shaped patterns on the mire image (visible at 0$^{\circ}$ and 180$^{\circ}$ in figure~\ref{fig:attachment}e).
This caused a break in the mires, and loss of mire data points in the v-shaped regions.
Our image enhancement algorithm interpolated to fill the lost mire data points. We experimented with a variety of support designs: 
(a) step-wise vs straight support,
(b) number of supports (two, three, four), and
(c) support thickness (1mm, 1.5mm, 2mm).
We found step-wise support to be unstable and fragile, and increasing the number of supports and/or its thickness resulted in higher loss of mire data points. 
Reducing the support thickness helped with more data points, however 1mm and 1.5mm support resulted in an unstable wobbly conical structure.
We found two diametrically opposite 2mm straight supports to work best.

\subsubsection{LEDs} Based on an experiment with 5 subjects, we decided on the number of LEDs to place in the illumination layer. 
We tested with 7, 10, and 16 LEDs (Figure~\ref{fig:attachment}a).
We found that seven LEDs led to dimly illuminated mires negatively impacting the accuracy of the image processing pipeline, and sixteen LEDs led to smudging of central mires (formed from rings closer to the LEDs) due to strong illumination and color bleeding. Thus, our final setup consisted of 10 LEDs placed at the outer boundary in a ring pattern (similar to Figure~\ref{fig:attachment}b red ring). Also, a butter paper was enveloped on the conical head, that acts as a light diffuser to make the empty regions of the placido uniformly lit.

\subsubsection{Simulation Environment}
\begin{figure*}[!t]
\begin{center}
    \centering
    \includegraphics[width=1.0\linewidth]{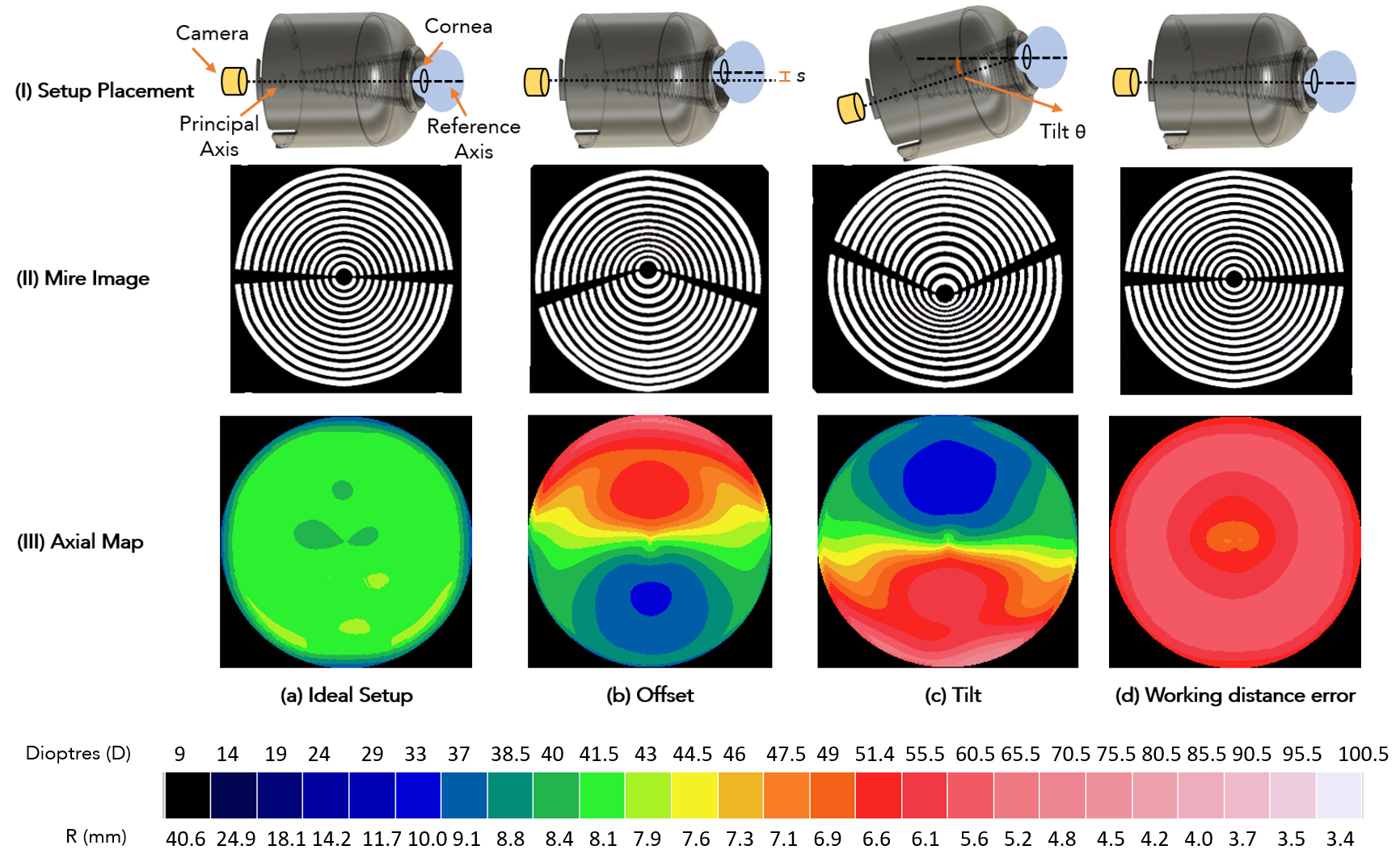}
\end{center}
\vspace{-4mm}
    \caption{Misalignment challenges during data collection. Top row shows the placement of our setup, middle row shows the captured mires, and bottom row shows the generated axial maps. The images were generated using our simulation setup with a perfect sphere imitating a cornea. The four columns show data collection with (a) ideal setup, (b) offset between reference and principal axes, (c) a tilt between the axes, and (d) wrong estimate of the working distance.
    } 
    \label{fig:challenges}
    \vspace{-4.0mm}
\end{figure*}

In order to test different placido designs, we setup a simulation environment in Blender~\cite{blender}, which is an open-source 3D computer graphics software.
Our simulated setup comprised of a placido attachment (by importing its .stl file), a diffused light source, a camera with parameters same as that of our data collection smartphone camera, and a reflection surface imitating the cornea.
The placido design and/or reflection surface were varied to obtain mire image as output (\textit{e.g.}, Figure~\ref{fig:challenges}II).
The generated image resolution is same as the cropped portion (Figure~\ref{fig:process_pipe}c) obtained from the smartphone's image, which we use as input to our image processing pipeline. 
The simulation environment helped us verify the working of our analysis pipeline in an ideal setting, and also enabled quick iterations over the attachment design. Moreover, it allowed us to quantitatively measure the sensitivity of our pipeline on various sources of error during data capture in the real world, which helped inform our design decisions.
\subsection{Challenges}
\label{sec:challenges}

We followed an iterative design process for \app{}---creating our placido attachment, building the data collection app, and developing the image processing pipeline. Each iteration consisted of collecting a small set of data (from $\sim$5 patients) and analysing it. During these iterations, we uncovered several problems which we solved with novel design decisions. Here, we focus on three core challenges that we identified.

\subsubsection{Offset}
This challenge is related to the misalignment between the camera's principal axis and cornea's reference axis (Figure~\ref{fig:challenges}).
The \textit{principal axis} passes from the center of the camera through the center of the placido rings, while the \textit{reference axis} is the symmetric rotational axis passing through the apex of the cornea~\cite{sinjab_2012} (shown as dotted and dashed line respectively, in Figure~\ref{fig:challenges}, row I). Ideally, the reference axis should pass through the center of the placido rings, such that both the principal axis and the reference axis are aligned (as in Figure~\ref{fig:challenges} I(a)).
An \textit{offset} between the two axes (shown as $s$ in Figure~\ref{fig:challenges} I(b)) results in distorted mires on the cornea (Figure~\ref{fig:challenges} II(b)), leading to incorrect diagnosis. 
The mire image  (Figure~\ref{fig:challenges} II(b)) was generated using our simulation setup emulating an eye as a sphere of radius 7.8mm, with the offset configuration of $s=4$mm. Since the placido rings are closer to the upper portion of the cornea and farther from the lower portion, the mires are spaced close-by in the upper region and farther apart in the lower.
This gives an incorrect perception of high and low curvatures in the two regions (shown as red and blue regions in Figure~\ref{fig:challenges} III(b)).
From our simulation setup, we found the topography computation to be quite sensitive to offset errors: a small offset of 3mm, 4mm and 5mm resulted in 5.5\%, 12.7\% and 14.1\% reconstruction error, respectively, on a 7.8mm radius spherical surface (also see Figure~\ref{fig:reconst_error}a).
Reconstruction error here is defined as the difference between the ground truth surface and the corneal surface generated by our image processing pipeline (an example of a generated corneal surface is shown in Figure~\ref{fig:process_pipe}g).
Thus, to achieve high accuracy, we need to ensure that the placido rings and the cornea center are in alignment with minimal/no offset.
To ensure this, we added a virtual crosshair (Figure~\ref{fig:process_pipe}b) \imwutadd{and real-time image quality checks} in our data collection app to help guide the data collector in minimizing offset.

\subsubsection{Tilt}
\begin{figure*}[!tb]
\begin{center}
    \centering
    \includegraphics[width=0.95\linewidth]{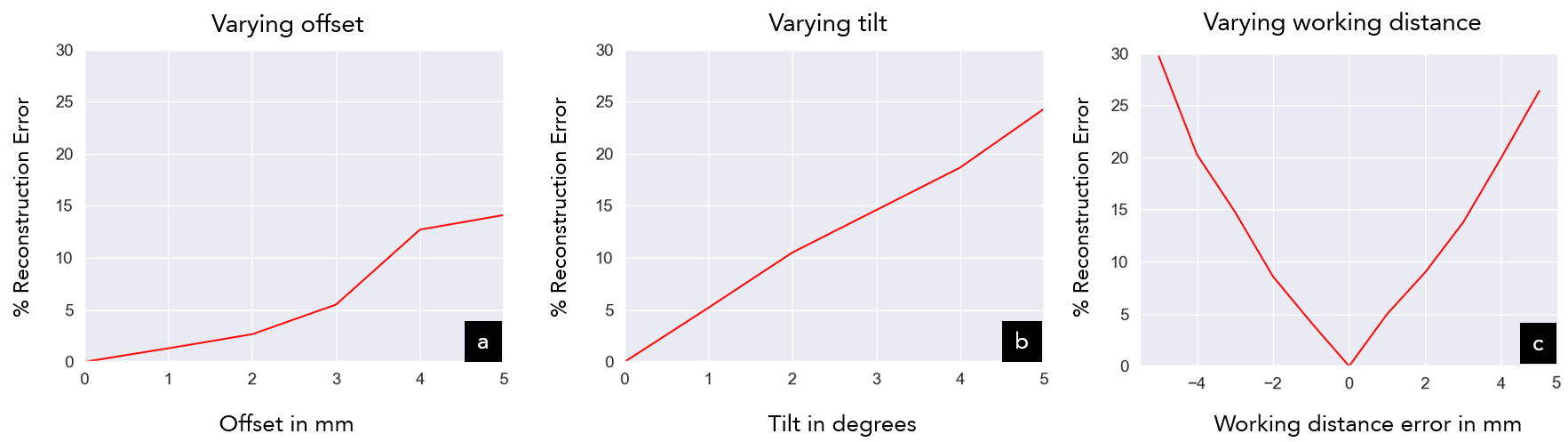}
\end{center}
\vspace{-4mm}
    \caption{Mean percentage reconstruction error for varying (a) offset, (b) tilt and (c) working distance error.} 
    \label{fig:reconst_error}
\vspace{-2mm}
\end{figure*}

The two axes can be at an angle ($\theta$ in Figure~\ref{fig:challenges} I(c)) leading to another source of misalignment.
As \app{} is a portable device, it is susceptible to tilt-based misalignment if the placido attachment is not completely touching the region around the patient's eye.
For instance, if only the lower portion of the placido attachment touches the patient's face, while the upper portion of the attachment does not, it will result in the principal axis to be at an angle with respect to the reference axis (Figure~\ref{fig:challenges} I(c)).
As rings in the lower portion will come closer to the eye compared to rings in the upper portion, it will lead to distorted mires (Figure~\ref{fig:challenges} II(c)) and incorrect curvature map (Figure~\ref{fig:challenges} III(c)). Such a tilt can result due to the variation in the structure of eye socket across individuals, and/or the data collector's lack of training in placing the device properly.
From our simulations, we found that a tilt angle of 3, 4, and 5 degrees resulted in 14.7\%, 18.6\% and 24.3\% reconstruction errors, respectively (also see Figure~\ref{fig:reconst_error}b).
To avoid such tilt misalignment and resulting errors, the placido attachment should have a snug fit around the eye socket (the contour of the orbital cavity), such that the front part of the attachment is fully touching the face. Apart from providing instructions to the data collector, we used the two supports on the placido attachment as a way to identify and minimise tilt. 

\subsubsection{Working Distance}
Even if the two axes are aligned, there can be another source of error. The distance between the smartphone camera and corneal apex, referred to as the \textit{working distance} (Figure~\ref{fig:attachment}a), is a crucial input to our image processing pipeline for computing the corneal curvature. Due to the difference in shape and size of the eye region (including orbital cavity) across patients, the working distance may vary, \textit{e.g.}, if a patient has a bulging eye, the working distance will decrease, while for a patient with a deep seated eye socket, the working distance will increase. Any error in working distance estimation results in incorrect topography computation. For instance, if the estimated working distance (70mm) is lower than the actual working distance (75mm), it will lead to the incorrect conclusion of high corneal curvature (Figure~\ref{fig:challenges} III(d)).
Our simulation analysis showed that the error in topography estimation is highly sensitive to the working distance estimation error---a working distance error of 1mm, 2mm and 3mm resulted in 5.0\%, 9.1\% and 13.8\% reconstruction error (also see Figure~\ref{fig:reconst_error}c), hence
an accurate estimate of working distance is essential for the success of a corneal topographer. Commercial topographers use special hardware to ensure a fixed working distance during image capture. Optikon Keratron uses a narrow depth-of-field camera such that the image is in focus only when the camera is within a fixed distance from the eye; Optikon Piccolo uses a laser beam; and Zeiss Humphrey Atlas uses a specialized placido attachment relying on parallax principle to compute working distance accurately. Smartphone cameras do not support a narrow depth-of-field, and due to low-cost constraints, we decided against using additional hardware. Instead, we relied on a novel computer vision based solution to estimate the working distance.
% \begin{table}[!htbp]
% \centering
%  \resizebox{\linewidth}{!}{
% %\small
% \begin{tabular}{ c | c c c c c c c c}
% \toprule
% \textbf{Parameter} & \textbf{Working Distance} & \textbf{Cone Length} & \textbf{Inner-Radius} & \textbf{Outer-Radius} & \textbf{Cone Semi-Vertical Angle} & \textbf{Mire Thickness} & \textbf{Focal Length} & \textbf{Sensor Dimensions}\\ 
% \hline
% \textbf{Value}  & 75.0 mm & 70.0 mm & 4.0 mm & 15.0 mm & 8.93$^{\circ}$ & 1.0 mm & 4.76 mm &  6.4 mm x 4.8 mm\\

%  \bottomrule
% \end{tabular}
% }
% \vspace{0mm}
% \caption{\textit{Optic specifications of the proposed system. Note: The Focal length and Sensor dimensions are given for the OnePlus 7T phone we have used for our experiments, but can be any other camera sensor.}}

% \label{tab:system-specs}
% \end{table}

\subsection{Data Collection App}
\label{sec:data_collection_app}

To collect data, we developed an Android application, \app{} app, which assists the data collector to capture good quality, \imwutadd{well aligned} and blur-free images.
For each patient, the data collector enters the age, sex and a hospital-assigned patient id. The patient id is the only link between the smartphone captured images, the clinical gold-standard Keratron data collected in the hospital, and the ground truth data of the actual keratoconus diagnosis by a senior ophthalmologist at the hospital. After entering the patient data, the app prompts the data collector to align the placido attachment over the patients' right eye and capture three images. The same is then repeated for the left eye.
To avoid any misalignment-related errors, the data collector has to ensure that there is neither \textit{offset} nor \textit{tilt} between the placido attachment's principal axis and the cornea's reference axis (refer Section~\ref{sec:challenges}).
 
To minimize offset, the \app{} app 
displays a crosshair on the camera preview, at the center of the placido attachment's projection using real-time Hough Circle~\cite{hough_circle} detection (see Figure~\ref{fig:process_pipe}b showing green-colored detected circle and red-colored crosshair). \imwutadd{It guides the data collector to align the center of the mire reflection with the red crosshair. Additionally, the app performs real-time checks on the incoming frames for exposure, sharpness and offset quality (refer Section \ref{sec:image-quality}). When all the checks pass, the camera preview displays a thick green border and automatically captures an image. Auto-capture ensures that the captured image is of good quality, by minimizing error due to hand movement that occurs during manual capture.} The data collector also asks the patient to look at the camera of the smartphone to help with this process.

To minimize tilt, the data collectors were trained to properly fit the placido attachment around the patient's eye socket. Moreover, a tilt-based misalignment can be identified from the reflection of the placido's support structure. Ideally, the support reflection should be horizontal at 0$^{\circ}$ and 180$^{\circ}$ angles in the mire image (Figure~\ref{fig:challenges} II(a)). A tilt (or even offset) results in the support reflection to be at an angle other than the horizontal angles (Figure~\ref{fig:challenges} II(c,b)). 
Even for keratoconus eyes, the support reflection will be mostly horizontal (maximum of $\pm$10$^{\circ}$, see Figure~\ref{fig:process_pipe}c, \ref{fig:qual_results}c.)
The data collector needs to adjust the position such that the support reflection becomes as close to horizontal as possible.
Although the support structure reduces the number of mire data points, it helped us in reducing the tilt-related misalignment error and is an important component of \app{}'s placido attachment.
Finally, to capture sharp mire images, the camera should focus on the corneal region. This is achieved by enabling auto-focus around the region centered at the crosshair. However, at times, the external mires were blurred, hence we also provided a tap-to-focus option to override the auto-focus region.

To ensure that at least one good quality image is captured for each eye, \imwutadd{as a secondary check,} the \app{} app prompts the data collector to review the images. After capturing three images for an eye, the app asks the data collector to approve/disapprove each image based on its \imwutremove{sharpness,}\imwutadd{horizontal tilt} alignment and corneal coverage\imwutadd{, which are not being checked automatically}. \imwutadd{The approve option is disabled in case any of the image quality checks fail, forcing the data collector to reject that image. Images which failed the quality check were still shown to the data collector as a feedback mechanism.}
As soon as the data collector approves one image from the set, we abort this review process for the corresponding eye. If none of the three images for an eye are approved, the data collector is prompted to capture three new images of the same eye. This process is repeated until at least one good image of each eye is captured. All the captured images are stored, irrespective of the data collector's approval. For each patient, the app stores a metadata file containing the patient's demographic information, data collector's input on image quality, app version, \textit{etc}. The collected data is auto-synced to a cloud-based storage in a best effort manner.

\imwutremove{Despite taking utmost care during data collection, 
we found a few captured images with significant tilt and/or offset misalignments. This may be attributed to COVID-19 as patients were wearing face masks, which often interfered with the placement of our device. Moreover, the pandemic prohibited in-person training of the data collectors. Thus, as a data quality check, we manually removed such misaligned eyes from our final dataset, based on the reflection of the support structure in the mire image.}

\subsection{Image Quality Checker}\label{sec:image-quality}
\imwutadd{
Capturing good quality mire images is challenging in a portable low-cost setting with minimal external hardware. Sources of error, such as tilt, offset and blur, can significantly impact the \app{} generated output maps. Thus, to ensure high quality image capture, we added an automatic quality checker module to the \app{} app.
The module checks for three sources of error---exposure, sharpness and offset---in real-time on the smartphone to give instantaneous feedback to the data collector and assist with data capture. We also perform an offline broken mires check on the captured images. Due to the compute requirement, this check could not be performed in real-time.
}

\noindent
\subsubsection{Exposure Check}
\imwutadd{
To ensure that the captured images are neither over- nor under-exposed, the \app{} app computes a histogram of the input frame's \textit{L channel} in the LAB color space. A frame is flagged as under-exposed if the histogram's maximum value is <125, and over-exposed if >20\% of the pixels are over 200 (similar to \cite{rdtscan}).}

\noindent
\subsubsection{Sharpness Check}
\imwutadd{To accurately segment and process mires, the captured image should be sharp and in focus. For ensuring sharpness, our app computes the variance of the edge intensity of the input frame after applying the Laplacian operator (as in \cite{rdtscan}). 
The test frame is considered sharp if its edge variance is >80\% of that of a good candidate image.}

\noindent
\subsubsection{Offset Check}
\imwutadd{Offset is the misalignment between the principal axis of the camera and the reference axis of the cornea (see Figure \ref{fig:challenges} I(b)). This can be computed from the captured frame as the translation between the placido attachment center and the mires center. Hough-circle routine~\cite{hough_circle} with different min-max radius thresholds is used to detect the placido attachment center and the mire center from the edge map of each frame. Using the camera parameters and the estimated working distance, the actual 3D location of the centers can be then obtained. A threshold is applied to only allow frames where the euclidean distance between the two centers is <1.0mm.}

\noindent
\subsubsection{Broken Mires Check}
\imwutadd{Broken/missing mires may arise due to eyelashes creating breaks in the mires or the eye not fully open (see Figure \ref{fig:failure-cases}(b)). In such cases the above three checks---exposure, sharpness and offset---may pass, however the broken mires can lead to inaccurate segmentation and missing data points, thus producing erroneous output heatmaps. To identify this, we compute the number of mires along each meridian using the \textit{radial mire scanning algorithm} (Section \ref{mire-placido}). If >5\% of the meridians have <24 mires (a fully open eye has at least 24 mires), the image is discarded.}

\imwutadd{The image quality checker module ensures that for each eye, the operator captures at least one good quality image passing all these checks.
It drastically improved the quality of captured images. For instance, adding the auto-capture and real-time quality checks reduced the mean offset error to 0.63$\pm$0.43 mm (on 101 eyes final dataset) from 1.59$\pm$0.86 mm (on 48 eyes pilot dataset).} 

\section{\app{} Image Analysis Pipeline}
Our offline analysis pipeline (Figure~\ref{fig:process_pipe}) takes as input the image collected via our setup, and outputs the corneal topography heatmaps along with quantitative sim-K values.
To compute the corneal topography from the input image, four sets of data points are needed: (1) camera parameters, (2) pixel location of the mires on the cornea, (3) 3D location of the placido rings with respect to the cornea, and (4) the working distance.
Only the camera parameters are known before hand from the smartphone specifications.

The image analysis pipeline extracts the mapping between the mires pixel location and corresponding placido rings. 
To compute the location of placido rings and working distance, we need to know two measurements---distance between the camera and base of the placido attachment (called $gap_{base}$), and distance between the corneal apex and top of the placido attachment (called $gap_{top}$) (Figure~\ref{fig:attachment}a). These fully determine the two sets of data points, as from the 3D print specifications, the distance of placido rings from the base of the attachment is known and working distance is the sum of $gap_{base}$, $gap_{top}$ and placido attachment length.
The four sets of data points are then combined using the Arc-Step method \cite{arc-step} and Zernike polynomial based surface fitting to generate the corneal topography. Below we describe each component of our analysis pipeline in detail:

% \begin{figure*}[!htbp]
\begin{figure*}
\begin{center}
    \centering
    \includegraphics[width=1.0\linewidth]{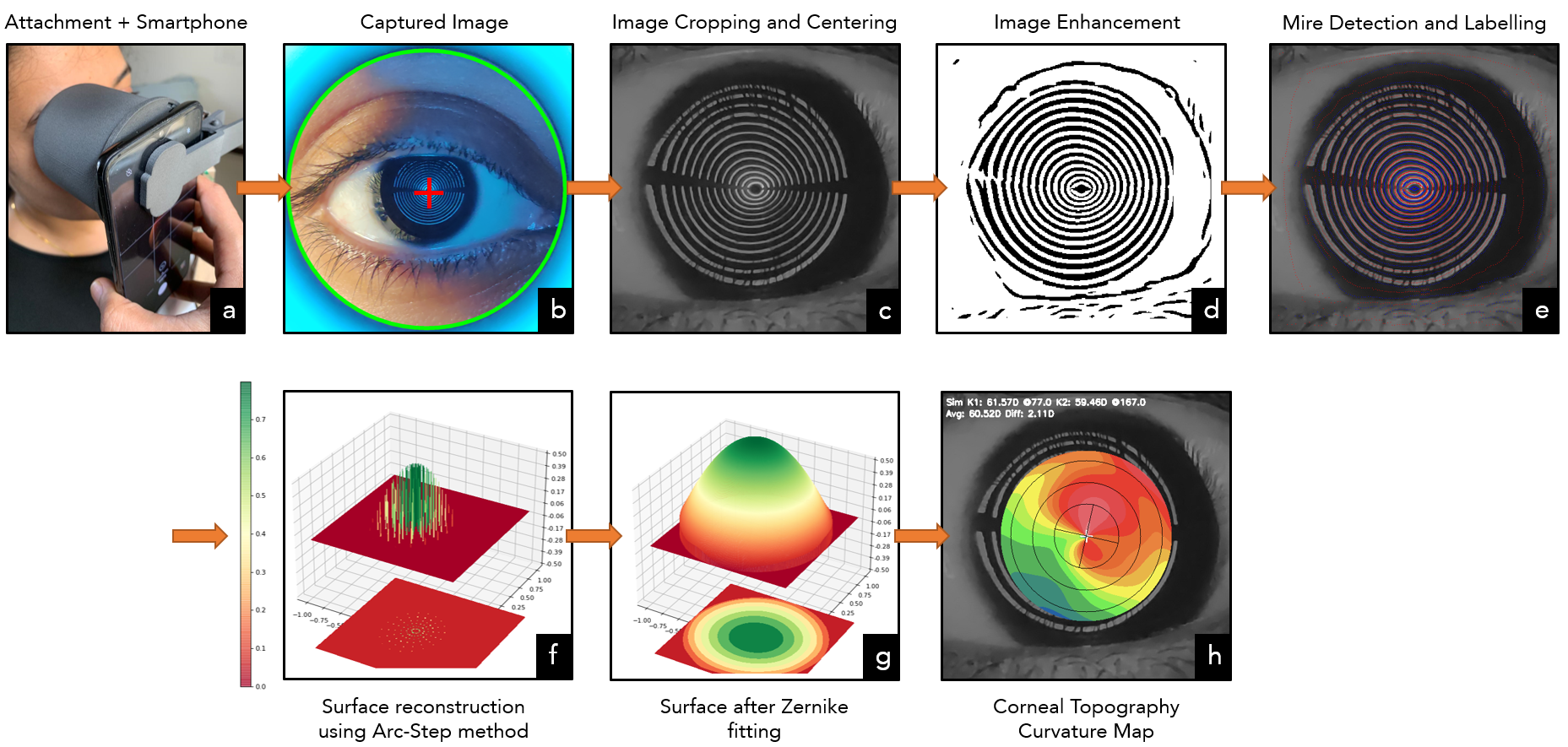}
\end{center}
\vspace{-4mm}
    \caption{\app{} image processing pipeline: (a) data collection process using \app{} app and placido attachment, (b) camera preview of patient's eye with red crosshair aligned with mire center, (c) center cropped gray-scale image, (d) enhanced image (e) image after mire detection (f) surface reconstruction using Arc-Step method, (g) corneal surface after Zernike fitting, (h) output topography map.} 
    \label{fig:process_pipe}
    \vspace{-4.0mm}
\end{figure*}

\subsection{Pre-processing: Image Cropping, Mire Segmentation, and Center Detection}
The input image, $I_C$, has a resolution of 3000 x 4000 pixels (width x height) and captures interior parts of the placido attachment along with the eye (Figure~\ref{fig:process_pipe}b). Since our placido setup is fixed, the corneal region of interest is always at the central portion of the image, hence we simply crop the central 500 x 500 pixels region (Figure~\ref{fig:process_pipe}c). Next, we convert the cropped image to gray-scale based on the {CCIR-601} standards to obtain, $I_G$ (Figure \ref{fig:process_pipe}c), as follows:
$I_G = (0.299\ I^R_C + 0.587\ I^G_C + 0.114\ I^B_C )$
where $I^R, I^G, I^B$ are the red, green and blue color channels of $I_C$, respectively.

The next step is to identify and segment out the mire pixels. First, we perform noise reduction on $I_G$ using Gaussian smoothening with a 7x7 kernel size. This also removes any spurious patches due to tear film irregularities or specular reflection, and is crucial to avoid false edge detection during segmentation.
Next, we normalize the noise reduced image for zero mean and unit standard deviation, followed by an algorithm to enhance the mire reflections (Figure \ref{fig:process_pipe}d). Since the mires are structurally similar to typical fingerprints, we adapted a fingerprint detection algorithm~\cite{finger_enhance} for mire enhancement. 
The enhancement algorithm estimates local orientation by computing gradients, and applies oriented Gabor filters to enhance the mires.
The enhanced image is then segmented
%Then the enhanced image is segmented 
using binary thresholding with a fixed constant obtained empirically, to obtain the binary mire image (Figure \ref{fig:process_pipe}d).

Finally, we estimate the center of the cornea region. Since the mire reflections are close to circular, we use Hough Transform~\cite{hough_transform, hough_circle} on an edge map to detect the smallest mire. The center of this circle is chosen as the center
% $(C_x, C_y)$ 
(red dot in Figure~\ref{fig:mire_detection}a). Although our method is quite robust, in rare cases where this center detection algorithm fails due to tear film distortion, \textit{etc.}, we allow for manual selection of the center, similar to Keratron.

\subsection{Mire-Placido Mapping}\label{mire-placido}
\begin{figure*}
\begin{center}
    \centering
    \includegraphics[width=1.0\linewidth]{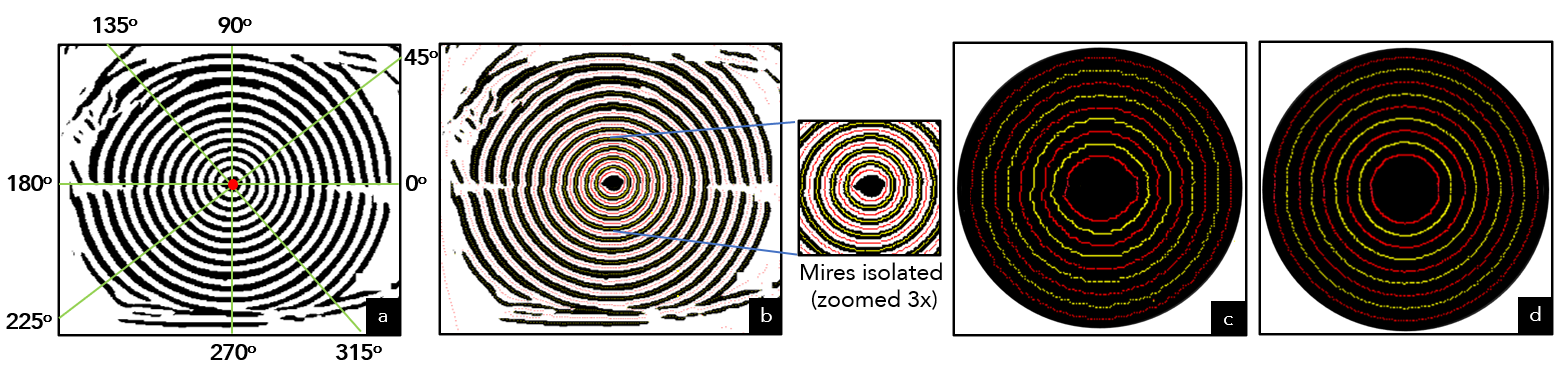}
\end{center}
\vspace{-2mm}
    \caption{(a) Illustration of the \textit{radial mire scanning algorithm}  with 8 rays separated by $45^{\circ}$ each, (b) detected mire points with alternating indices in red / yellow color. Effect of anti-aliasing scheme: (c) mire points before anti-aliasing, and (d) after anti-aliasing.
    } 
    \label{fig:mire_detection}
\vspace{-2mm}
\end{figure*}

In this step, we map each mire to the corresponding placido ring from which it was reflected.
To robustly identify mire location and number them, we propose a \textit{radial mire scanning algorithm}, which scans for mires along radial rays starting from the cornea center moving radially outwards. We discretize the angular space into $k$ angles and shoot a ray for each angle. Figure \ref{fig:mire_detection}a shows an example with 8 rays separated by $45^{\circ}$ each. At a given angle, the algorithm moves along the ray while keeping track of the black and white segments it encounters. For any black/white segment, the centroid of pixel locations corresponding to this segment is the computed mire location.
We empirically found that taking centroid helps in reducing the noise and provides a more accurate estimate of the mire location, compared to a scheme which uses the boundary points of the segments as mire location.
The mire scanning process is repeated for each of the $k$ angles ($k=360$) to obtain mire pixel locations along with their respective indices. Finally, we apply median filtering on the radial co-ordinate of points on the \textit{same} mire to remove any outliers. The output mire points are depicted in Figure~\ref{fig:mire_detection}b, with alternating indices in red and yellow color.

\subsubsection{Anti-Aliasing} 
Since the cornea region of interest is only the central region, our input image resolution is low (500x500 pixels). This combined with the thresholding step for mire segmentation introduces aliasing artifacts in the binary image (Figure~\ref{fig:process_pipe}d), which in turn leads to axis-aligned step aliasing in the detected mires (Figure~\ref{fig:mire_detection}c).
We notice that the image before thresholding, Figure~\ref{fig:process_pipe}c, has no \textit{visible} aliasing artifacts, mainly because pixels covered partially by a mire have intensity values proportional to their coverage, thus giving a smooth appearance. 
Based on this principle, we leverage the intensity values of pixels to anti-alias the detected mires. Instead of simple averaging to compute the centroid of each segment, we perform an intensity value based weighted average of the pixel locations.
This resulted in a more accurate mire location estimate with no/minimal aliasing (Figure~\ref{fig:mire_detection}d).

\subsection{Placido Ring Location and Working Distance Estimation}
\label{sec:working_distance_estimation}

The 3D location of placido rings with respect to the cornea and the working distance are inputs to the Arc-Step method which estimates the corneal topography. To compute these two data points, 
we need to compute two distances---$gap_{base}$: distance between the camera and base of the placido attachment, and $gap_{top}$: distance between the corneal apex and top of the placido attachment (Figure~\ref{fig:attachment}a).
Since the location of placido rings from the base of the attachment is known (from the 3D print specifications),
in order to compute rings' location with respect to cornea, we need $gap_{top}$. Similarly, as $working\: distance = placido\:length + gap_{base} + gap_{top}$, both the distance measurements are required for computing working distance. As the placido attachment is rigidly fixed to the smartphone cover, $gap_{base}$ is constant across patients; however, due to the variation in shape and size of eye (including cornea and orbital cavity), $gap_{top}$ varies across patients, and is harder to compute accurately.
Estimating these distances accurately is crucial as the reconstruction error is sensitive to working distance (Section~\ref{sec:challenges}).

\subsubsection{Measuring camera-attachment distance ($gap_{base}$)}
In commercial topographers, $gap_{base}$ is known, as these devices are manufactured using precise industrial tools. However, in our case, this distance is hard to measure precisely, because (a) we clip the placido attachment to the smartphone, and (b) the distance between the camera casing and camera location inside the smartphone is unknown.
We use a calibration sphere of known radius (7.8mm) for estimating $gap_{base}$ (similar to \cite{phone-kt-3D-print}). We place the calibration sphere onto the top of our placido attachment (\textit{i.e.}, with $gap_{top}=0$), and capture a mire image.
As the $gap_{top}$ is known, we run our image analysis pipeline with varying values of $gap_{base}$, and choose the value that minimizes the error between the estimated radius of the calibration sphere and the actual radius. We computed $gap_{base}$ once for our setup, and keep it fixed throughout our evaluation.

\subsubsection{Measuring attachment-eye distance ($gap_{top}$)}
Once $gap_{base}$ is known, $gap_{top}$ can be estimated using two properties of the human eye: (1) the radius of the human cornea is similar across individuals (7.79$\pm$0.27mm), and (2) even in advanced keratoconus, the shape of the cornea is distorted mainly in the central region, while the outer regions remain unaffected~\cite{about-kt-2}. This implies that for a given $gap_{top}$, the radius of the mire reflections in the outer region will be approximately same across patients. Thus, a 1:1 mapping exists between $gap_{top}$ and radius of an outer mire. To learn this mapping, we use our simulation environment. We generated mire images on a sphere of radius 7.8mm (imitating cornea) at varying $gap_{top}$ (from -5 to +5mm), and compute the corresponding average radius of the $20^{th}$ mire. We used the $20^{th}$ mire as it is typically situated in the outer region of the cornea. We then fit a simple regression model ($y=a/x+b$) to estimate the mapping between the $20th$ mire radius (in pixels) and $gap_{top}$\footnote{$gap_{top}$ is usually a negative value since the top of the placido attachment touches the region around the patient's eye socket, making the corneal apex protrude inwards.} (in mm). 
Thus, for a patient's eye during analysis, we compute the average radius of the $20^{th}$ mire, and refer to the learnt mapping to estimate $gap_{top}$.
(Note: We cannot directly calculate the error in $gap_{top}$ estimation, as the ground truth $gap_{top}$ value is unknown.)

\begin{figure*}[!tb]
\begin{center}
    \centering
    \includegraphics[width=1.0\linewidth]{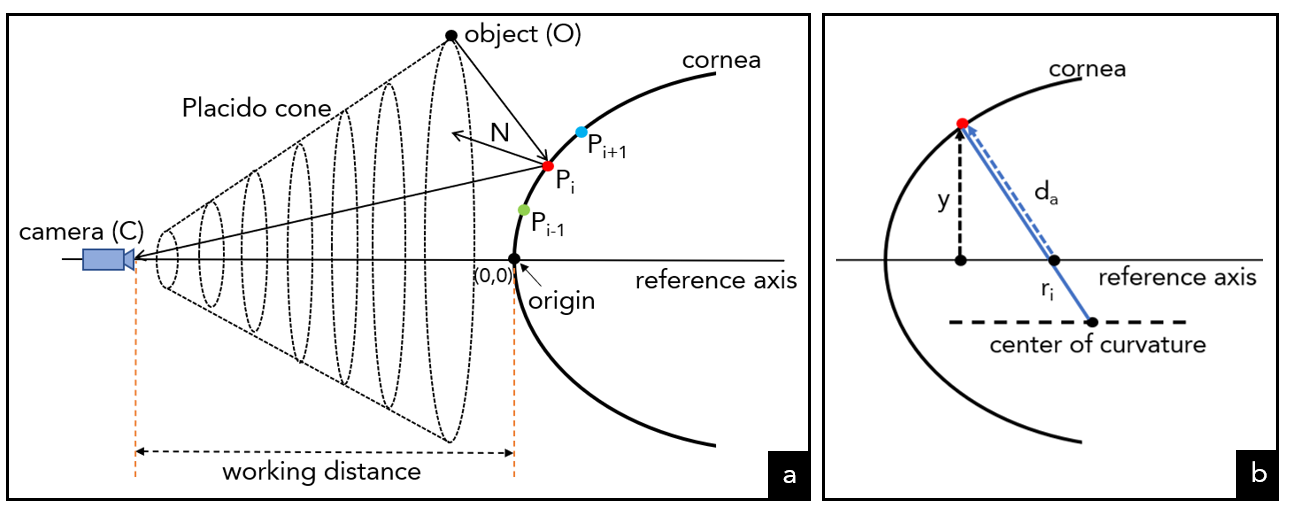}
\end{center}
\vspace{-4mm}
    \caption{(a) Arc-Step method illustration, (b) Tangential ($r_i$) and the axial ($d_a$) radius of curvature at a given point on the cornea.} 
    \label{fig:arc-step}
\end{figure*}

\subsection{Topography Estimation: Arc-Step Method and Zernike Polynomial Based Surface Fitting}
The camera parameters, pixel location of the mires along $k$(=360) radial rays, placido location with respect to the cornea, and working distance can be combined to reconstruct the 3D location of the corneal points, using the Arc-Step method~\cite{arc-step} (Figure~\ref{fig:process_pipe}f).
The Arc-Step method estimates the corneal elevation along meridians (corresponding to the $k$ radial rays) starting from the cornea center and emanating radially outwards. It is an iterative method, where each iteration progressively estimates the 3D segment of the cornea along a meridian between two consecutive mires, by fitting a cubic polynomial. Given the cubic between $P_{i-1}$ and $P_{i}$, the algorithm proceeds to fit a cubic between $P_{i}$ and $P_{i+1}$ (Figure~\ref{fig:arc-step}a). Two consecutive cubic segments (\textit{e.g.}, $(P_{i-1}, P_{i})$ and $(P_{i}, P_{i+1})$) are constrained to be smooth at the intersection point ($P_{i}$). A $C^2$ continuity condition is imposed, providing three constraints---matching position, first derivative, and second derivative. Since a cubic has four parameters, one more constraint is needed. 
For a given mire pixel location ($P_{i}$) and camera parameters, the ray from the camera center ($C$) to $P_{i}$ can be determined. This ray, in turn should be a reflection of the ray from the corresponding placido point ($O$) to the cornea. The 3D location of the placido point $O$ is known, we can combine this with the law of reflection (angle of incidence = angle of reflection) to obtain our final constraint for fitting the cubic. The Arc-Step method outputs a set of smooth curves along the $k$ meridians. 

To reconstruct the corneal surface, we fit an analytic surface using Zernike polynomials~\cite{zernike_1, zernike_2, zernike_3} (Figure \ref{fig:process_pipe}g). Zernike polynomial is known to effectively represent corneal surfaces, given sufficient number of basis polynomials~\cite{forward_ray_tracing}. We experimented with polynomials up to degree 15, 
which corresponds to 66 Zernike basis polynomials, 
and found degree 8 (= 45 basis polynomials) to be a satisfactory fit with a low reconstruction error.
This surface representation of the cornea can be used to generate the axial and tangential maps (Figure \ref{fig:process_pipe}h), which clinicians use to diagnose keratoconus.

\subsection{Corneal Topography Heatmaps and Quantitative Values}\label{sec:heatmaps}
The reconstructed corneal surface is represented by dioptric power heatmaps, which help doctors visualize the corneal shape for diagnosis.
The most commonly used topographic maps are the axial and tangential map. These maps plot the dioptric power, $D = \frac{\mu-1}{r}$, where $\mu = 1.3375$ is the refractive index of the corneal fluid and $r$ is the radius of curvature (either axial or tangential) in meters.

\subsubsection{Axial/Sagittal Map} The axial radius of curvature of a curve, at a given point, is the distance, $d_a$, from the point to the reference axis measured along the normal at that point~\cite{tan_to_axial_klein} (Figure~\ref{fig:arc-step}b). Axial maps provide an abstract overview, and are useful in screening of corneal abnormalities, especially keratoconus diagnosis~\cite{sinjab_2012}.

\subsubsection{Tangential/Instantaneous Map} The tangential radius of curvature of a curve, at a given point, is the radius of the circle $r_i$ which best fits the curve at that point locally (Figure~\ref{fig:arc-step}b). The tangential maps\footnote{Keratron refers to tangential map as `curvature map', however `curvature' is a general term, hence we avoid using that for tangential map.} are more sensitive compared to axial maps and depict finer details. Thus they are used for localization of defects in the cornea~\cite{sinjab_2012}.

\noindent (Note: For ease of explanation, we have described these curvatures in terms of one-dimensional curves. Please refer~\cite{arizona_curvature} for a complete explanation and derivation.)

\subsubsection{Simulated keratometry (sim-K)} 
\updateremove{With respect to quantitative values, we estimate the simulated keratometry (sim-K) values, which measures the dioptric power of the steepest (sim-K1) and flattest (sim-K2) meridians in the central 5mm diameter region of the cornea.}\updateadd{With respect to quantitative values, we estimate the sim-K values~\cite{metrices} --- (a) sim-K2: by measuring the average dioptric power of the flattest meridian in the central 3mm diameter region of the cornea, and (b) sim-K1: by computing the average dioptric power at $90^o$ from the flattest meridian to obtain the steepest meridian.}

In clinical practice, ophthalmologists refer to the axial and tangential heatmaps generated by commercial topographers.
They look for patterns in the heatmaps, such as asymmetric bow tie, superior steep, and butterfly~\cite{sinjab_2012}, along with the quantitative values---sim-K1 and sim-K2---to identify keratoconus and its severity.
\section{Data Collection}\label{sec:data-collection}
To evaluate our proposed approach, we collected eye data using \app{} and Keratron for the same set of patients at a local eye hospital. 
All details of our study were approved by the hospital's Institutional Review Board.

\subsection{Method}
The data was collected during \imwutremove{Dec'20-Jan'21 and Mar-Apr'21}\imwutadd{Jun-Aug'21}
%in India,
in collaboration with a local eye hospital.
The eye hospital is one of the leading tertiary eye care and teaching institution in Bengaluru, India.
It has >15 eye doctors (including >3 cornea specialists), >10 optometrist, and treats approximately 500+ patients every day (pre-COVID). The hospital has several state-of-the-art equipment, including two corneal topographers--Optikon Keratron and WaveLight Oculyzer (similar to Oculus Pentacam).
Both the corneal topographers were placed in the same room, and were operated by a staff optometrist (female, 25-40 years age) who had received training in operating both these devices.
For our data collection, one of the researchers trained the same staff optometrist to use the \app{} phone app.
As data was collected during the COVID-19 period, with intervals of restricted mobility, the training was conducted online.

The training mainly included using the \app{} phone app for capturing good quality \imwutremove{blur-free}eye images.
The researcher provided a brief overview of the research project to the data collector, and demonstrated how to use the app over a video call.
The data collector was asked to repeat the process with another staff member acting as patient. During training, the researcher provided prompt feedback on the images captured by the data collector.
After entering the patient id, sex and age on the \app{} phone app, the app prompts to align the attachment over the patients' right eye and \imwutadd{auto-}capture images till a good quality image for that eye is captured. The same process is repeated for the patients' left eye. 
To minimize offset and tilt related misalignment errors, the data collector was trained to (a) align the center of the mires with the red crosshair displayed on the camera preview, and (b) ensure that the support reflection is horizontal at 0 and 180 degrees.
As the placido attachment touches the patients' face (inside the contour of the orbital cavity), we asked the data collector to wipe the attachment's outer surface using alcohol swab, after every use.

While collecting data from the phone app, patients were instructed to blink a few times before the image capture, keep their eye wide open, look at the smartphone camera at the center of the concentric rings, and cover the other eye (the eye not being image captured) using their hand.
Several instructions were given to the data collector---correctly entering the patient id in the phone app, as that was the only link between the smartphone captured images and the Keratron data; ensuring that the LEDs are on while capturing images; tapping on the smartphone screen over the ring reflections for the camera to focus appropriately, if needed; and making certain that the phone is vertical such that it is neither leaning towards nor away from the patient.

We asked the staff optometrist to try recruiting every patient that any doctor referred for a corneal topography test.
The participants were given an IRB-approved consent form to read and sign before their data was collected.
The order of data collection was to conduct the Keratron test, 
followed by collecting data from the \app{} phone app.
Every evening, the data collector connected the phone to the hospital's WiFi network to auto-sync the collected images along with the metadata to a cloud-based storage.
Moreover, the data collector also uploaded anonymized Keratron data (labeled with the same patient id) comprising of axial map, tangential map, and placido ring image, for each eye of the patient, on another cloud-based storage. 
The ground truth data for each eye of the patient was the actual diagnosis by a senior ophthalmologist at the hospital, and consisted of the eye's refractive error and keratoconus classification; this was shared with us at the end of each data collection cycle.

The \app{} data was collected using a OnePlus 7T smartphone, utilizing its main camera (48 MP, f/1.6 aperture, 4.76mm focal length, 6.4mm x 4.8mm sensor size) for capturing images. All the data was collected in the same room having the Keratron device, by placing two chairs---one each for the data collector and patient---facing each other. No lighting related changes were made to the room, as the placido attachment blocks any external light. Participants were not paid to participate in the study, and spend 5-10 minutes for data collection from both the devices.

\subsection{Participants and Dataset}
\imwutadd{57} patients (\imwutadd{35} female, \imwutadd{22} male) with an average age of \imwutadd{23.7}$\pm$\imwutadd{7.6} years (\imwutadd{12}-\imwutadd{51} years age range) participated in the data collection.
(Note: Due to COVID-19, the number of patients visiting hospital dropped drastically, hence we were able to collect a \imwutadd{relatively} small dataset in a 2-months time period.)
After \imwutadd{auto-}removing faulty images captured using \app{} (with \imwutremove{blur, tilt, and/or center alignment}\imwutadd{broken mires}), the final dataset comprised of \imwutadd{101} distinct eyes, with \imwutadd{67} non-keratoconus and \imwutadd{34} keratoconus eyes.
The Keratron dataset consisted of three images (placido rings reflection image, axial map, and tangential map) and three quantitative values (sim-K1, sim-K2 and PPK) for each eye; the \app{} dataset has demography data (age, sex), three or more images for each eye captured using the app, and metadata tagging a \imwutremove{blur-free}\imwutadd{good quality} image for each eye.
As measured by Keratron, for the non-keratoconus and keratoconus eyes, the mean sim-K1 value was \imwutadd{45.6}$\pm$\imwutadd{2.1}D and \imwutadd{51.1}$\pm$\imwutadd{4.9}D, the mean sim-K2 value was \imwutadd{44.6}$\pm$\imwutadd{2.0}D and \imwutadd{46.8}$\pm$\imwutadd{3.3}D, and the mean PPK value was \imwutadd{0.05}$\pm$\imwutadd{0.14} and \imwutadd{0.84}$\pm$\imwutadd{0.33}, respectively.
The ground truth data for each eye was the actual diagnosis by a senior ophthalmologist (cornea specialist) at the hospital, comprising of the patient's refractive error (spherical, cylindrical, and axis) and keratoconus classification (non-keratoconus, keratoconus).
For the diagnosis, the ophthalmologist combines manual analysis (using slit lamp and retinoscope), and results from Keratron and Oculyzer devices.

\imwutadd{Apart from this dataset, we had also conducted a pilot data collection phase during Dec'20-Jan'21 and Mar-Apr'21 period, without the real-time on-device image quality checker module in the app. The pilot data had 36 patients. After offline analysis, we found that only 48 of the 72 eyes had at least a single good quality image. Of the discarded images, 14 eyes had a large offset error (more than 1mm), 2 eyes had bad exposure or blur, and 8 eyes had broken mires. As we had to remove 33.3\% of the collected eyes, we decided to not use this pilot data for final analysis.}
\section{Result} \label{sec:result}
To evaluate the effectiveness of \app{}, we used the simulated environment (Section~\ref{sec:placido_attachment}) to quantitatively evaluate the accuracy and correctness of our image analysis pipeline in an ideal setting. We then conducted a comparative evaluation of the corneal topography maps generated by \app{} and Keratron, by \updateremove{two}\updateadd{four} experienced doctors. The study included \imwutadd{101} distinct eyes: \imwutadd{67} without keratoconus and \imwutadd{34} with keratoconus. We also compare the quantitative sim-K1 and sim-K2 values computed by the two devices.
Finally, we compute the repeatability of measurements reported by \app{} on 10 randomly sampled eyes, with two distinct images for each eye.

\subsection{Simulation Study}
We set up a simulated environment for our system, as described in Section \ref{sec:placido_attachment}. 
We experimented with three different types of artificial corneal surfaces: (1) a spherical surface of radius 8mm, (2) an ellipsoidal surface with 8mm, 10mm, 12mm lengths of the three principal semi-axes, and (3) a set of five Zernike surfaces obtained from the corneal topography of our patient data. 
The projected image generated from the simulation setup (of resolution 500x500 pixels) is passed through our analysis pipeline to reconstruct the surface.
Comparing the artificial corneal surface with the surface reconstructed from our pipeline, we found the mean percentage reconstruction error to be 0.07\%, 0.04\% and 2.41\% and the error in radius of curvature to be 1.24\%, 0.95\% and 3.25\%, for the spherical, ellipsoidal and Zernike surfaces, respectively. The errors
% (Table \ref{tab:sim-error}) 
for the three surfaces are minimal, thus demonstrating the high measurement accuracy of our analysis pipeline under ideal conditions.

\begin{table}[!tb]
\centering
\caption{\imwutadd{Precision/Recall and Sensitivity/Specificity for \app{} and Keratron from the doctors' evaluation.\\
%*: after removing retake, $^{\dagger{}}$: assuming if either eye (left/right) of a patient was marked as `keratoconus' both the eyes were evaluated as having `keratoconus'
}}
\resizebox{\linewidth}{!}{
\small
\imwutadd{
\begin{tabular}{| c | c | c | c c c | c c |}
\toprule
\textbf{Examiner} & \textbf{System} & \textbf{\# of eyes}* & \textbf{Precision} & \textbf{Recall} & \textbf{F1 Score} & \textbf{Sensitivity} & \textbf{Specificity}\\ 
\hline
PPK  Metric & Keratron & 101 & 0.91 & 0.94 & 0.93 & 94.1\% & 95.5\%\\
\hline
\multirow{2}{*}{Ophthalmologist \#1} & Keratron & 73 & 0.69 & 1.00 & 0.82 & 100.0\% & 65.8\% \\
 & SmartKC & 86 & 0.94 & 0.94 & 0.94 & 93.5\% & 95.9\%\\
\hline
\multirow{2}{*}{Ophthalmologist \#2} & Keratron & 61 & 0.93 & 1.00 & 0.96 & 100.0\% & 93.3\% \\
 & SmartKC & 88 & 1.00 & 0.92 & 0.96 & 92.3\% & 100.0\%\\
%Ophthalmologist \#2 & SmartKC & 68 & 0.72 & 0.96 & 0.83 & 96.3\% & 73.2\%\\ This was before SimK correction
\hline
\multirow{2}{*}{Ophthalmologist \#3} & Keratron & 91 & 0.54 & 0.94 & 0.69 & 94.1\% & 52.6\% \\
& SmartKC & 83 & 1.00 & 0.94 & 0.97 & 93.7\% & 100.0\%\\
\hline
\multirow{2}{*}{Ophthalmologist \#4} & Keratron & 84 & 0.57 & 1.00 & 0.73 & 100.0\% & 53.8\% \\
& SmartKC & 99 & 1.00 & 0.94 & 0.97 & 94.1\% & 100.0\%\\
\hline
\multirow{2}{*}{Overall} & Keratron & 82 & 0.68 & 1.00 & 0.81 & 100.0\% & 64.5\% \\
& SmartKC & 97 & 1.00 & 0.94 & 0.97 & 94.1\% & 100.0\%\\
 \bottomrule
\end{tabular}
}}
% \vspace{-4mm}

\label{tab:precision_recall}
%\vspace{-6mm}
\end{table}

% \app{} achieves a combined Sensitivity / Specificity of 85.7\% and 78.6\% as compared to gold standard device Keratron's Senstivity / Specificity of 100.0\% and 78.6\%, demonstrating \app{}'s efficacy in a clinical-setting.
\subsection{Patient Study}
\begin{figure*}[!t]
\begin{center}
    \centering
    \includegraphics[width=1.0\linewidth]{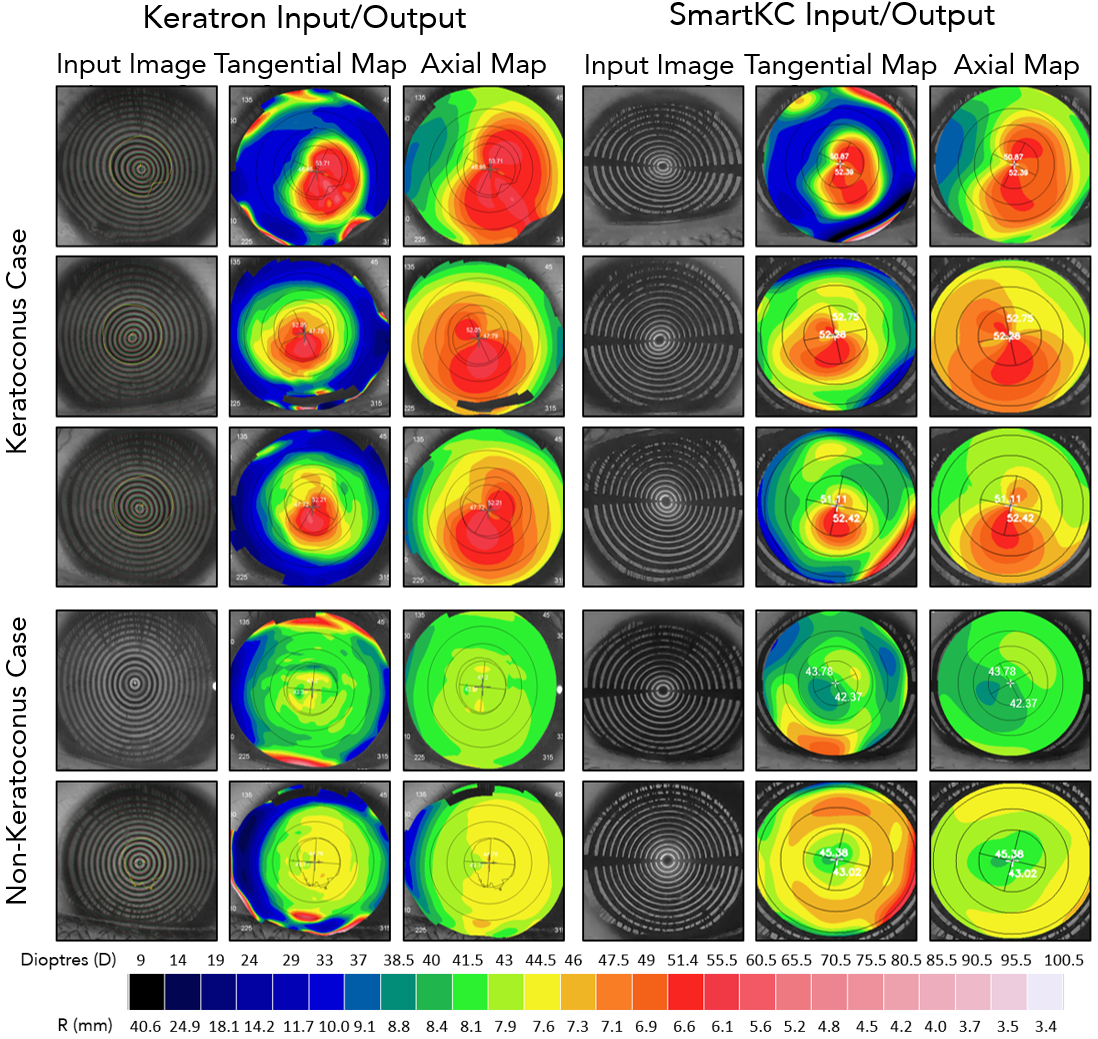}
\end{center}
\vspace{-4mm}
    \caption{The axial maps generated by Keratron and \app{} are qualitatively very similar. \textbf{(Rows 1-3)} show the input image, tangential map and axial map for keratoconus eyes, and \textbf{(rows 4-5)} for non-keratoconus eyes.}
    \label{fig:qual_results}
\vspace{-4mm}
\end{figure*}

In clinical practice, ophthalmologists refer to the axial and tangential corneal topography maps, along with the sim-K1, sim-K2 and PPK values, to diagnose keratoconus.
We used the same color palette as Keratron to generate our axial and tangential maps. We also overlaid sim-K1 and sim-K2 values at the center of the heatmaps, similar to Keratron generated maps (Figure \ref{fig:qual_results}). To evaluate the clinical value of \app{} in a real-world setting, we asked \updateremove{two}\updateadd{four} \updateremove{senior}ophthalmologists (cornea specialists, 1 male and \updateremove{1}\updateadd{3} female, working at Sankara Eye Hospital\footnote{\href{https://sankaraeye.com/}{https://sankaraeye.com/}}, Bengaluru, India) to rate \imwutadd{202} pairs of output corneal topography maps, \imwutadd{101} each from our \app{} system and from the gold-standard Keratron device.
Each question comprised of a pair of axial and tangential maps for the same eye, from one of the devices (\app{} or Keratron). 
The images were randomized to minimize bias across severity of keratoconus and across the two devices. The doctors were asked to choose one of the three options: non-keratoconus, keratoconus, and retake (if the heatmaps were unusual due to misalignments and/or tear film disturbances). The doctors did not receive any extra training or instructions for this task.

\updateadd{Keratoconus is usually bilateral, \textit{i.e.}, it affects both the eyes~\cite{bilateral}. Even in our dataset, all the 17 patients had keratoconus in both their eyes. Leveraging this phenomenon, for our evaluation, if either eye (left/right) of a patient was marked as `keratoconus' by a doctor, we labelled both their eyes having `keratoconus'.} After removing the images labeled as `retake', we computed the precision/recall and specificity/sensitivity scores for each ophthalmologist individually as well as \updateremove{combined}\updateadd{for the overall doctor's panel} (Table~\ref{tab:precision_recall}). The ophthalmologist ratings were compared with the actual ground truth diagnosis that we collected separately.
\updateremove{The \textit{combined scores} were calculated as follows: (a) if both the ophthalmologists agreed on a label for a question, we used that label, (b) if one ophthalmologist label it as retake, we used the other ophthalmologist's label, and (c) if the ophthalmologists reported conflicting labels (except retake), we used the worst case scenario (\textit{i.e.}, if the labels are `non-keratoconous' and `keratoconus', we used `keratoconus' as the final combined label).}
\updateadd{The overall scores of the four doctors was computed as a simple majority decision ($\geq50\%$ agreement)~\cite{jama-diabetic-retino}. If no majority was reached, the sample was marked as `retake'. The inter-rater agreement computed for the non-retake images across the four doctors in terms of Cohen's Kappa score was 0.86 for the \app{} images and  0.98 for the Keratron images.}

On evaluating the overall scores, \app{} achieved a sensitivity of \imwutadd{94.1}\% and specificity of \imwutadd{100.0}\% as compared to Keratron's sensitivity of \imwutadd{100.0}\% and specificity of \imwutadd{64.5}\% (Table~\ref{tab:precision_recall}).
Though \app{}'s sensitivity is lower than Keratron, it is in the acceptable >\updateremove{85}\updateadd{90}\% range~\cite{posterior_good}.
\updateremove{Note that \app{} accounted for 4 false-negatives---2 were suspect keratoconus (that were neither identified by Keratron PPK-value, nor by doctors who rated Keratron heatmaps as `retake') and the other 2 were mild keratoconus cases. None of the moderate/severe keratoconus cases were missed by \app{}. The panel of doctors reached a majority decision for 96.0\% of the \app{} images, and 81.2\% of the Keratron images.}
\updateadd{Note that \app{}'s lower sensitivity was due to two false negatives that were suspect keratoconus cases, which were neither identified by Keratron PPK-value, nor by doctors (who rated those two Keratron heatmaps as `retake'). None of the moderate/severe keratoconus cases were missed by \app{}.}

\updateremove{Keratoconus is usually bilateral, \textit{i.e.}, it affects both eye. Even in our dataset, all the 17 patients had keratoconus in both the eyes. Leveraging this phenomenon, we computed \textit{combined$^{\dagger{}}$ scores}, which is similar to the \textit{combined scores} except that if either eye (left/right) of a patient was marked as `keratoconus' both the eyes were evaluated as having `keratoconus'. This improved the sensitivity of \app{} to 93.5\%, while Keratron's sensitivity remained unchanged.}

\updateadd{Interestingly, we found the precision for Keratron to be significantly lower than that of \app{}. On analyzing the doctors' ratings, we found that the majority of false positives for Keratron (72\%) were due to incorrect working distance estimation by Keratron, which results in heatmaps with artificially high curvature (\textit{e.g.}, Figure~\ref{fig:challenges} III(d)). 
In contrast, our device \app{} can robustly estimate the working distance accurately (refer Section \ref{sec:working_distance_estimation}), thus handling such cases well, which resulted in its high precision.}
\updateremove{We found both Keratron and \app{} to be low on precision.
This means both the devices identified keratoconus cases correctly; however, there were false positives, \textit{i.e.}, non-keratoconus cases were marked as keratoconus.
In a medical setting, it is critical to not miss any keratoconus case, so a low precision hints that our doctors were conservative in their rating.}
We also noticed that \updateremove{21}\updateadd{19} Keraton and \updateremove{6}\updateadd{4} \app{} heatmaps were rated retake by the \updateremove{two doctors}\updateadd{panel of doctors} in the \updateremove{combined}\updateadd{overall} evaluation (Table~\ref{tab:precision_recall}). On further analysis, we found that the retake Keratron heatmaps had mainly tear film disturbances~\cite{sinjab_2012}, while \updateremove{incorrect working distance estimate}\updateadd{tilt/offset errors during data collection or tear film disturbances} resulted in ``\textit{unusual}''/``\textit{atypical}'' heatmaps from \app{}. 

\begin{figure*}[!tb]
\begin{center}
    \centering
    \includegraphics[width=1.0\linewidth]{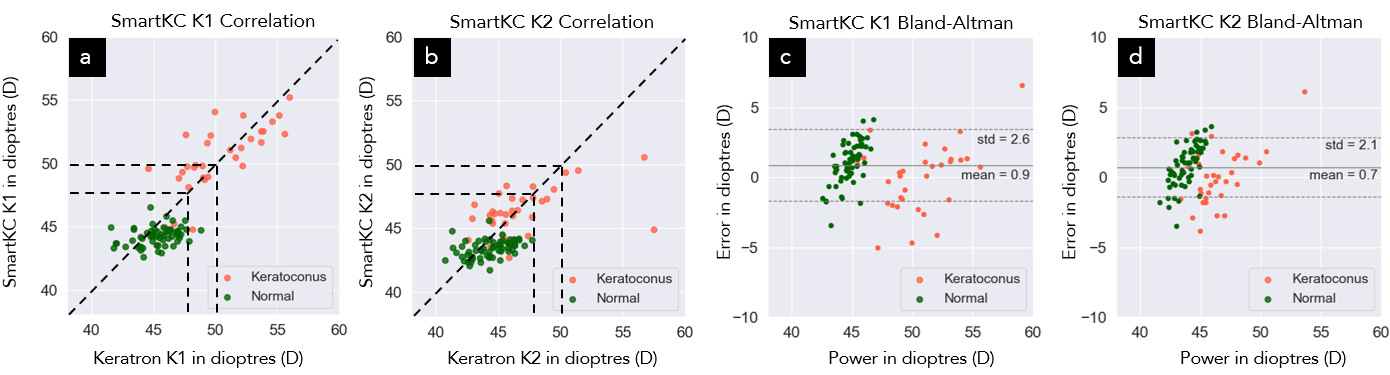}
\end{center}
\vspace{-4mm}
    \caption{(a), (b): Correlation plots and (c), (d): Bland-Altman plots for \app{} versus Keratron generated sim-K1, sim-K2 values.} 
    \label{fig:corr_plots}
\end{figure*}

\begin{figure*}[!tb]
\begin{center}
    \centering
    \includegraphics[width=1.0\linewidth]{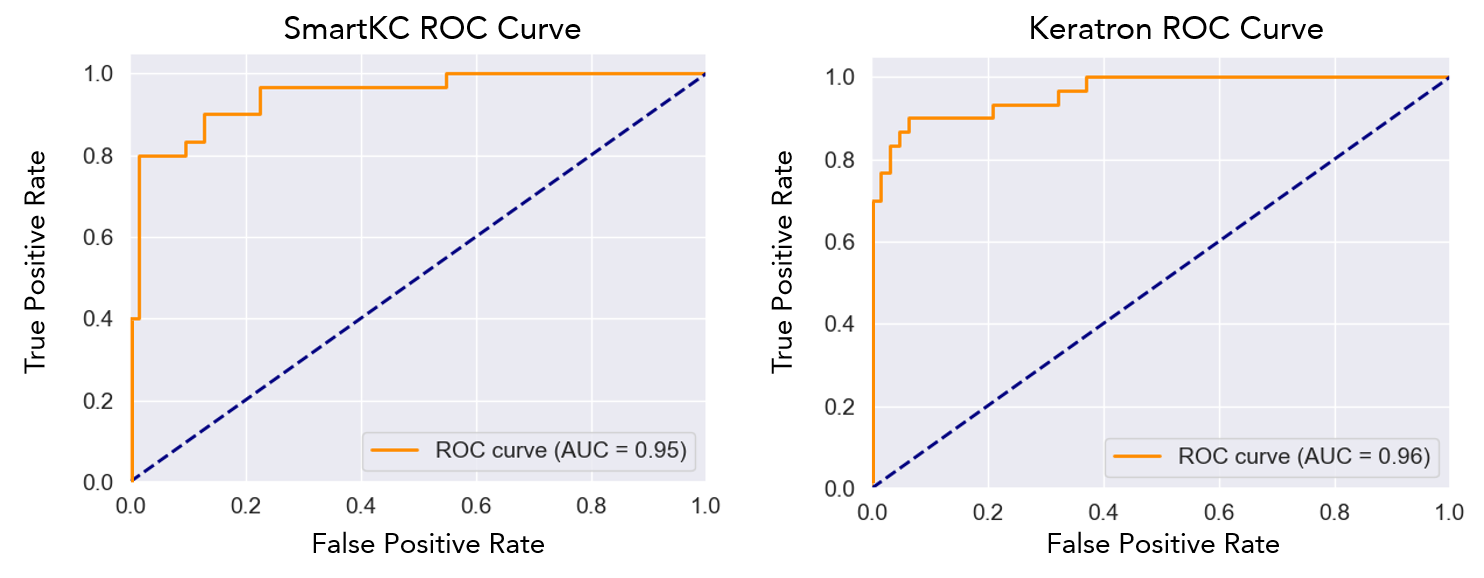}
\end{center}
\vspace{-4mm}
    \caption{ROC curves showing efficacy of \app{} and Keratron as a screening device, using the thumb rule for sim-K1 and sim-K2 values based automatic diagnosis.}
    \label{fig:roc_curve}
\end{figure*}

Moreover, we analyzed the simulated keratometry values, by comparing the sim-K1 and sim-K2 values obtained from \app{} with the values reported by Keratron.
As shown in Figures~\ref{fig:corr_plots}a,b, the values were strongly correlated, with sim-K1 had Pearson's r(\imwutadd{99}) = \imwutadd{0.78}, p < 0.01, and sim-K2 had Pearson's r(\imwutadd{99}) = \imwutadd{0.62}, p < 0.01. Bland-Altman plots (Figure~\ref{fig:corr_plots}c,d) of the same measurements highlight that the mean difference between \app{} and Keratron for sim-K1 was $\imwutadd{0.9}\pm\imwutadd{2.6}$D and for sim-K2 was $\imwutadd{0.7}\pm\imwutadd{2.1}$D. This shows good agreement between the sim-K1, sim-K2 values generated by \app{} and Keratron. The sim-K1 and sim-K2 values can also be used to screen keratoconus. As a rule of thumb, either of the sim-K1 or sim-K2 values below 47D implies non-keratoconus, 47-50D implies mild keratoconus, 50-54D implies moderate keratoconus, and above 54D implies severe keratoconus.
Figure \ref{fig:roc_curve} shows the ROC curves for \app{} and Keratron as a classifier using the sim-K1/K2 thumb rule to automatically diagnose keratoconus. The area under the ROC curve (AUC) is \imwutadd{0.95} for \app{} and \imwutadd{0.96} for Keratron.
Though these numbers are very promising, we found ophthalmologists to rely on axial and tangential heatmaps for diagnosis, over sim-K1, sim-K2 values.

Keratron also outputs a PPK (percent probability of keratoconus) \cite{CLMI} value in green, orange or red color representing no, mild or severe keratoconus, respectively. We found that Keratron's PPK value based classification achieved a sensitivity of 94.1\% and specificity of 95.5\% (Table~\ref{tab:precision_recall})\updateremove{, and is not perfect}. \updateadd{We plan to include a similar metric in our SmartKC system for automatic classification of keratoconus in future.}

Qualitatively the captured mire images and output topography heatmaps generated by \app{} were similar to those of Keratron. Figure~\ref{fig:qual_results} presents \imwutadd{five} sample results. Each row consists of two input/output (input mire image and output maps) sets for Keratron and \app{}, from the same patient's eye. The \imwutadd{first three} rows illustrate the outputs for keratoconus eyes, and the \imwutadd{last two rows} for non-keratoconus eyes.
These results clearly indicate the efficacy of our proposed \app{} system under real-world settings.

\subsection{Repeatability Experiment}
For reliability, the output generated by a corneal topographer should be consistent across measurements. To assess that the topography maps and sim-K values generated by \app{} are consistent across different measurements, we evaluated the repeatability of measurements on 10 distinct eyes. These 10 eyes were randomly sampled from our dataset where we had at least two good quality images for the same eye. We found the mean absolute difference in the reconstructed corneal surface in the central 7mm region was $0.003 \pm 0.002$mm ($2.41\% \pm 0.96\%$), and the mean sim-K1 and sim-K2 difference was $0.47 \pm 0.45$ D and $0.36 \pm 0.29$ D, respectively.
On performing a similar analysis on Keratron, on 10 random images, we found the mean sim-K1 and sim-K2 difference to be -0.21 $\pm$ 0.22 D and -0.13 $\pm$ 0.27 D, respectively.
The results show that \app{} is reliable and the diagnosis is repeatable.

\subsection{Usability Feedback}
We conducted a feedback session with the data collector after the completion of our study. The data collector found the placido attachment convenient to handle, the \app{} app easy to use, and appreciated the crosshair, tap-to-focus \imwutadd{and auto-click} functionalities.
She mentioned that some patients were hesitant that \app{} might hurt their eye socket. To reduce their apprehensions, ``\textit{I gave the device to such patients, and asked them to touch the device's orbit by their hand, and even touch it around their face and eyes. After they were sure that the device has smooth curved edges and won't hurt, they agreed to participate}''.
None of the patients complained about the intensity of the LED lights in \app{}.
Compared to Keratron, the data collector found \app{} to be ``\textit{faster}'', as Keratron require capturing four images for each eye with manual adjustment of the device (to find the eye center and to keep Keratron at the correct distance from the eye), along with manually verifying the quality of each image in the computer attached to the device. This switching between Keratron and computer for capturing and quality checking, respectively, was cumbersome. As \app{} was light-weight, it was easier to maneuver, and having input/output in the same device simplified the data capture process.

\section{Discussion} \label{sec:discussion}

In this paper, we aim to develop a low-cost, portable, smartphone-based corneal topography system. \app{} combines a 3D printed placido attachment, an LED light array, and a\imwutadd{n intelligent} smartphone app to robustly capture placido reflections on the cornea.
Our image processing pipeline analyzes the captured image via novel approaches for extracting pixel location of mires using the radial mire scanning algorithm, and for computing the working distance using $gap_{base}$ and $gap_{top}$. The set of computed data is combined using Arc-Step method with Zernike polynomials based surface fitting to construct the corneal surface. This surface is used to generate axial and tangential heatmaps, along with sim-K1 and sim-K2 estimates.
Our evaluation showed a high correlation between the gold-standard Keratron sim-K and our predicted sim-K values, \imwutremove{with a mean difference of $0.2\pm3.0$ D}\imwutadd{with a Pearson correlation coefficient of 0.78}. We also achieved an acceptable sensitivity of \imwutadd{94.1}\% and specificity of \imwutadd{100.0}\% for keratoconus diagnosis by \updateremove{two}\updateadd{four} ophthalmologists based on heatmaps generated from \app{}. We believe that \app{} has the potential to be an important tool in detection and progress tracking of keratoconus, especially in remote areas. Although our results look promising, more research is needed to make the system more robust and at par with gold-standard medical devices.

\subsection{Applications}
Corneal topography is an important tool for diagnosis and monitoring of several corneal diseases, such as keratoconus, corneal marginal degeneration, keratoglobus, and post-LASIK ectasia. Moreover, it is an important input while performing eye surgeries, including refractive error correction surgery. Although in this paper we focused on keratoconus, the same system can be extended to other applications.

\subsubsection{Mass screening for keratoconus} Based on our discussions with ophthalmologists, a low-cost portable corneal topographer can be very useful for mass-screening drives, especially in low-resource regions. For example, school teachers can be trained to use \app{}, and screen students every quarter for early diagnosis of keratoconus. Similarly, community health workers can carry the \app{} attachment to perform screening of keratoconus in remote areas effectively.

\subsubsection{Corneal Astigmatism} Astigmatism is a condition where the eye has different refractive power along different axes. The most common kind of astigmatism is corneal astigmatism, which occurs due to slight deformations of the cornea causing it to be curved more steeply in a specific direction. The sim-K values computed by \app{} provide an estimate of the principal curvatures of the corneal surface. We believe that the difference of these values can estimate the cylindrical power, and the direction along which sim-K1 is computed can estimate the cylindrical axis. Our initial analysis in this direction looks promising, however we need to collect more data of patients with astigmatism for evaluation.

\subsubsection{Dry Eye} The images captured by \app{} can be analysed for estimating the height of tear between the iris and lower eyelid, called tear meniscus height. A tear meniscus height of less than 0.25mm is indicative of the dry eye disease. A single device for diagnosis of multiple eye conditions can prove to be very useful.

\subsection{Affordable Cost}
To keep \app{} low-cost, we relied on minimal hardware: a 3D-printed clip-on placido attachment (\$23.50), a USB-connected array of 10 LED lights (\$6.50), and a type-C to USB 3.0 female adapter (\$3), totaling to \$33. This cost will further reduce, if we produce/buy these parts in bulk. The cost of the OnePlus7T smartphone used in our experiments is \$450, which makes the total cost of the setup \$483. In comparison, commercial devices cost significantly more,\textit{ e.g.}, Keratron costs 20x more at \$10,000. Although we tested our setup with only a OnePlus 7T smartphone, \app{} can be used with any Android smartphone with a camera resolution of 12MP or more. Moreover, since our target audience, such as community health workers and teachers, already carry a smartphone, the incremental cost of the setup (<\$33) is very affordable.
Our main design decisions to enable a low-cost topographer included: (a) building a support-based placido ring structure, instead of an expensive multi-color cone with opaque and transparent rings, (b) using an LED array directly powered by the smartphone which alleviated the need for extra circuitry, battery and casing (unlike~\cite{phone-kt-3D-print}),
(c) relying completely on smartphone's default camera (prior works add extra micro lens for higher image quality~\cite{phone-kt-3D-print, bullseye}) enabled by image enhancement methods \imwutadd{and image quality checks}, and (d) proposing a novel software-only approach for estimation of working distance (unlike Optikon Piccolo, Zeiss Atlas, that use extra hardware for this).
Other than reducing cost, our minimalist design approach resulted in a light-weight attachment of 140 grams. 

\subsection{Diagnostic Output}
Many metrics have been proposed for fully automatic diagnosis of keratoconus based on the topography heatmaps, such as PPK and aCLMI~\cite{CLMI}. However, based on our discussions with three senior ophthalmologists, these quantitative metrics are not perfect. Edge keratoconus cases, for example, may not adhere to pre-defined thresholds on these metrics. Moreover, these metrics do not localise the deformed corneal region, which is useful for treatment and explanation purposes. Hence, doctors primarily rely on the topography heatmaps to obtain a holistic view of the corneal surface. This was the main reason for us to focus on generation of the topography heatmaps which can be directly interpreted by doctors without any additional training.

We believe that a fully automated diagnosis can be helpful in cases where a trained ophthalmologist is not available for interpreting the heatmaps, especially in resource constrained regions. Medical devices, such as Keratron report PPK (percentage probability of keratoconus) values, which provide a diagnosis of keratoconus severity: PPK<0.20 is no keratoconus, 0.20<PPK<0.45 is mild keratoconus, and PPK>0.45 is severe keratoconus. \updateremove{Although PPK is not a perfect metric (Table 2), it can serve as an alternative when an ophthalmologist is not available.}\updateadd{In our evaluation we found that Keratron’s PPK value based classification achieved a sensitivity of 94.1\% and specificity of 95.5\% (Table \ref{tab:precision_recall}).} In future, we aim to compute PPK or similar metrics based on our heatmaps. We plan to apply deep learning approaches on the heatmaps to automatically predict the keratoconus severity \imwutadd{and location}. Such learning based approaches can even combine raw mire images, heatmaps, and computed sim-K values to give high accuracy outputs.

\begin{figure*}[!tb]
\begin{center}
    \centering
    \includegraphics[width=1.0\linewidth]{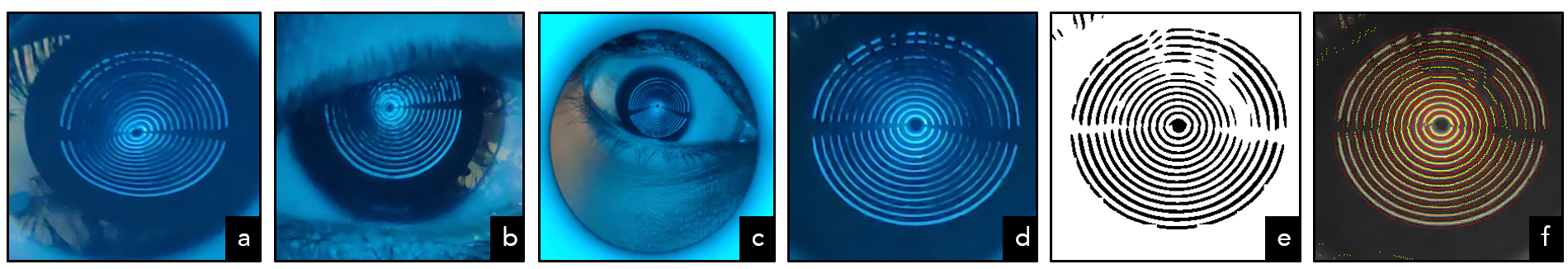}
\end{center}
\vspace{-4mm}
    \caption{Examples of discarded images, due to (a) external reflection, (b) eye not fully open \imwutadd{or broken mires}, (c) rotated eye, (d) blur, which lead to failure in (e) mire segmentation and (f) mire-placido mapping.} 
    \label{fig:failure-cases}
    \vspace{-4.0mm}
\end{figure*}

\subsection{Discarded Images}
Commercial topographers rely on chin- and head-rest for capturing good quality images. In contrast, \app{} is hand-held, which may result in blurry images and offset/tilt-based misalignment errors.
As a result, \imwutadd{in the pilot data collection,} \imwutremove{from}\imwutadd{out of} the 72 eyes images captured from 36 patients, we could only use 48 eyes for evaluation. The discarded images had errors such as (see Figure~\ref{fig:failure-cases}): (a) external reflections, (b) partially open eye, (c) rotated eye, and/or (d) image blur, which lead to failure in (e) mire segmentation and (f) mire-placido mapping. \imwutremove{To improve the quality of collected data, we plan to improve our data collector training module with specific instructions and checklists, and incorporate real-time feedback mechanisms which can help data collectors with the capture process. 
Moreover, such offset and tilt checks can be automated, thus capturing images automatically when these conditions are satisfied.}\imwutadd{To improve the quality of data collection, we added an image quality checker module to the \app{} app that provides real-time feedback to the operator. This module significantly improved the captured image quality. In the final dataset, only 11.4\% eyes were discarded (due to the broken mires offline check) as compared to 33.3\% eyes discarded in the pilot study. The broken mires check is currently performed offline due to compute limitations, however in future we wish to incorporate this in an updated version of the phone app.}

\imwutadd{Other factors that might have contributed to the poor image quality: (a) patients wearing masks due to the COVID-19 pandemic which interfered with the placement of placido attachment, (b) restricted mobility due to which researchers could not provide any in-person training to the data collector, and (c) learning curve in using \app{}.
We believe that these factors will further reduce in future, resulting in improved data quality, and thus improved accuracy.}

\subsection{Limitations}
We acknowledge \imwutremove{two key}\imwutadd{a few} limitations of this work.
First, due to the COVID-19 lockdowns, the number of (keratoconus) patients visiting the hospital reduced. Moreover, a few patients were hesitant in using an additional contact device. As a result, we were able to collect data for only \imwutremove{36}\imwutadd{57} patients. \imwutremove{This along with the image quality problems, further reduced our dataset.}Although, the current evaluation seems promising and many prior mobile health sensing work has $\sim$50 subjects~\cite{seismo, hemaapp, spirosmart}, we believe an evaluation with a bigger sample size, and with patients from different demography is needed before real-world deployment. \imwutadd{Second, our data collection was limited to a single smartphone device. We would like to evaluate \app{} on multiple devices, although we expect devices with a minimum camera specification of 12 MP to perform similarly.}
Third, we found the topography computation to be sensitive to errors in the working distance estimation.
Even a 1mm error resulted in 5\% reconstruction error, which may lead to unusual heatmaps with atypical color patterns.
In future, we plan to improve the working distance estimation algorithm, along with exploring low-cost hardware alternatives to further improve its accuracy.
\imwutadd{Fourth, despite the real-time feedback provided by the \app{} app, $\sim$11\% of the samples had to be rejected due to the broken mires offline check. We plan to implement a simple version of the broken mires checker algorithm on the smartphone. Finally, the SmartKC app was used only by one data collector. For wider deployment, we need to further study and improve the usability of the app.}
\section{Conclusion}

Low-cost mobile corneal topographers can be an important tool for early detection and mitigation of keratoconus, a leading cause of blindness among younger demographics in low- and middle-income countries. In this paper, we presented \app{}, a smartphone-based keratoconus diagnosis system based on the placido disc principle. We evaluated our system on \imwutadd{101} distinct eyes, and achieved a sensitivity of \updateremove{87.8}\updateadd{94.1}\% and a specificity of \updateremove{80.4}\updateadd{100.0}\% for diagnosis of keratoconus by ophthalmologists based on topography heatmaps and sim-K values generated by \app{}. The evaluation also showed a high correlation between the gold-standard Keratron's sim-K values and our predicted sim-Ks, with a \imwutremove{mean difference of 0.2 $\pm$ 3.0 D}\imwutadd{Pearson correlation coefficient of 0.78}. An automatic classification of keratoconus based on sim-K values yielded an AUC of \imwutadd{0.95}. Although more research is needed to bring \app{} at par with commercial medical devices, we believe that \app{} has the potential to be used as a keratoconus screening device, especially in resource-constrained areas.

\begin{acks}
We would like to thank Bill Thies, Muthian Sivathanu, Alex Mariakakis, and Pratyush Kumar for providing constructive feedback on the paper, Swamy for helping with the data collection, and all the participants for their time and patience.
\end{acks}

% \clearpage
\bibliographystyle{format/ACM-Reference-Format}
\bibliography{references}

%\appendix
\end{document}